# Scenario-based Tsunami hazard assessment for Northeastern Adriatic coasts

Antonella Peresan[1] and Hany M. Hassan[1,2]

[1]National Institute of Oceanography and Applied Geophysics- OGS, Trieste, Italy
[2]National Research Institute of Astronomy and Geophysics, 11421 Helwan, Cairo, Egypt

## Abstract

Significant tsunamis in Northern Adriatic are rare and only a few historical events were reported in the literature, with sources mostly located along with central and southern parts of the Adriatic coasts. Recently, a tsunami alert system has been established for the whole Mediterranean area; however, a detailed description of the potential impact of tsunami waves on coastal areas is still missing for several sites. This study aims at modelling the hazard associated with possible tsunamis, generated by offshore earthquakes, with the purpose of contributing to tsunami risk assessment for selected urban areas located along the Northeastern Adriatic coasts. Tsunami modelling is performed by the NAMI DANCE software, which allows accounting for seismic source properties, variable bathymetry, and non-linear effects in wave's propagation. Preliminary hazard scenarios at the shoreline are developed for the coastal areas of Northeastern Italy and at selected cities (namely Trieste, Monfalcone, Lignano and Grado). A wide set of potential tsunamigenic sources of tectonic origin, located in three distance ranges (namely at Adriatic-wide, regional and local scales), are considered for the modelling; sources are defined according to available literature, which includes catalogues of historical tsunami and existing active faults databases. Accordingly, a preliminary set of tsunami-related parameters and maps are obtained (e.g. maximum run-up, arrival times, synthetic mareograms), relevant towards planning mitigation actions at the selected sites.

## 1. Introduction

The coastal areas of Adriatic Sea are exposed to a range of natural hazards, particularly floods, wind storms, droughts, earthquakes, and tsunamis.Although strong tsunamigenic earthquakes are not frequent in the Adriatic Sea,considering the coastal exposure and vulnerability their impacts could be considerable. Indeed, even a moderate tsunami may have a relevant effect on some areas due to the presence of sites of high historical/cultural/touristic interest (e.g. the city of Venice) and fragile ecosystems (e.g. lagoons and river deltas). Defining hazard scenarios for a wide range of possible tsunamigenic earthquakes is of essential importance to reduce their potential socioeconomic impacts on coastal communities and set up sustainable development plans for inland and offshore areas. The adequate estimation of expected hazard, along with the characterization of exposed assets and their vulnerability, provides the basis for tsunami risk assessment.

Historical records indicate that the Adriatic region experienced considerable impacts from earthquakes and tsunamis in the past (Maramai et al., 2014; 2019; 2021). For instance, the 30 July, 1627 Gargano earthquake (M6.7) and subsequent tsunami caused widespread destruction along the eastern Adriatic coast (Patacca and Scandone, 2004). More recently, the 15 April, 1979 Montenegro earthquake ($M_w$6.9) generated a local tsunami that affected coastal areas in Montenegro and Albania. Altogether, Pasarić et al. (2012) identified 27 tsunamigenic events that occurred during the last 600 years: nine of them are located on the west side of the Adriatic Sea, while eighteen are located along its eastern side (6 are classified as reliable, 10 as unreliable, while 2 are categorized as events of meteorological origin). In 2019, an updated tsunami catalogue for the Mediterranean region was published, namely the Euro-Mediterranean Tsunami Catalogue (EMTC), which includes a detailed survey of tsunamis of seismic origin in the Adriatic Sea (Maramai et al., 2019); this study identified the potential tsunamigenic events, including a level of uncertainty in the location and magnitude of seismic events.

Significant tsunamis are especially rare in Northern Adriatic, where only a few historical events were reported in the literature, with sources mostly located along central and southern parts of the Adriatic coasts (Maramai et al., 2007; Tiberti et al., 2008). Anyway, recent review studies, based on historical information, support relevant impacts of past tsunamis in the North East Adriatic region in the years 1348 and 1511 (Maramai et al., 2021). Despite the evidence of sporadic, but possibly damaging events, a detailed investigation of the potential impact of tsunamis of seismic origin on coastal activities and communities along the Northern Adriatic coasts is not yet available. Therefore, this work aims to provide different scenario based tsunami hazard measures along the shoreline for the area of interest (e.g. maximum wave amplitude, estimated time of arrivals) given the available source, bathymetry, and topography datasets. The results are essential toward a comprehensive understanding and assessment of exposure, vulnerability and risk, and thus to increase resilience of coastal communities in Northeastern Adriatic region. Moreover, these results are also needed to develop emergency plans, as well as to properly inform residents and visitors about the potential tsunami risk in the region and to convey adequate instructions in the event of a tsunami warning.

Quite recently, in fact, a tsunami early warning system has been established for the entire Northeast Atlantic, and the Mediterranean region, referred to as NEAMTWS (Tsunami Warning and Mitigation System in the Northeastern Atlantic, the Mediterranean and connected seas). In this framework, the CAT-INGV system(Centro Allerta Tsunami, https://cat.ingv.it/en/), managed by INGV, operates as a tsunami warning centre for the whole Mediterranean, including the Adriatic Sea (Amato et al., 2021). At the national level, CAT-INGV operates within the formal framework of the National Alert System for Earthquake Generated Tsunamis (SiAM - Sistema Allertamento Maremoti), established in 2017 (Directive of the President of the Council of Ministers, 17 February 2017) with the aim of disseminating alert messages to the territory, including local authorities. A debate is ongoing about the possibility of delivering *last-mile* messages to alert the population. In case a potentially tsunamigenic event occurs, the system provides alert messages expressed by the following three levels, in order of increasing severity: *"information"* (green), *"advisory"* (orange) and *"watch"* (red), based on the specific decision matrix (Fig. 1) reported in Amato et al. (2021). According to this matrix, the distance between a site of interest and the epicentre is divided into three possible categories: local (distance ≤ 100 km); regional (100 km < distance ≤ 400 km); and basin-wide (distance > 400 km). The "*watch*" level is defined when seismic information indicates that the coast may be hit by a tsunami with a wave amplitude greater than 0.5 m offshore (along 50 m isobaths), and/or when the tsunami run-up at the shoreline is expected to be greater than 1 m. The "*advisory*" level is indicated, instead, when the run-up is expected to be less than 1 m. Since 2017, several events have been located and assigned a watch alert level by the CAT-INGV system, including the Aegean Sea earthquake in 2017 (Mw=6.6), and the more recent events in Crete (2$^{nd}$May 2020), Aegean Sea (30$^{th}$October 2020), Northern Algeria (18$^{th}$ March 2021) and the Turkey earthquakes (6$^{th}$ February 2023).

<table>
<tr><th rowspan="2">Depth</th><th rowspan="2">M</th><th rowspan="2">Epicenter Location</th><th rowspan="2">Tsunami Potential</th><th colspan="3">ALERT LEVEL VS DISTANCE</th></tr>
<tr><th>Δeq ≤ 100 km</th><th>100 km < Δeq ≤ 400 km</th><th>Δeq > 400 km</th></tr>
<tr><td rowspan="6">< 100 km</td><td>5.5 ≤ M ≤ 6.0</td><td>Offshore or Inland ≤ 100 km</td><td>Nil</td><td colspan="3">Information Bulletin</td></tr>
<tr><td rowspan="2">6.0 < M ≤ 6.5</td><td>Inland (40 km < Inland ≤ 100 km)</td><td>Nil</td><td colspan="3">Information Bulletin</td></tr>
<tr><td>Offshore or near the coast ( Inland ≤ 40 km)</td><td>Potential of weak local tsunami Δeq < 100 km</td><td>LOCAL Tsunami ADVISORY</td><td colspan="2">Information</td></tr>
<tr><td>6.5 < M ≤ 7.0</td><td rowspan="3">Offshore or Inland ≤ 100 km</td><td>Potential of destructive local tsunami Δeq < 100 km | 400 km</td><td>LOCAL Tsunami WATCH</td><td>REGIONAL Tsunami ADVISORY</td><td>Information</td></tr>
<tr><td>7.0 < M ≤ 7.5</td><td>Potential of destructive regional tsunami Δeq < 400 km | basin</td><td colspan="2">REGIONAL Tsunami WATCH</td><td>BASIN-WIDE Tsunami ADVISORY</td></tr>
<tr><td>M > 7.5</td><td>Potential of destructive tsunami in the whole basin any Δeq</td><td colspan="3">BASIN-WIDE Tsunami WATCH</td></tr>
<tr><td>≥100 km</td><td>M ≥5.5</td><td>Offshore or Inland ≤100 km</td><td>Nil</td><td colspan="3">Information Bulletin</td></tr>
<tr><td>any</td><td>any</td><td>Inland > 100 km</td><td>Nil</td><td colspan="3">Nil</td></tr>
<tr><td colspan="4"></td><td>LOCAL</td><td>REGIONAL</td><td>BASIN-WIDE</td></tr>
</table>

Figure 1- CAT-INGV tsunami decision matrix (https://cat.ingv.it/en/tsunami-alert/alert-procedures/decision-matrix, last accessed May 2023).

A still open challenge is to go beyond the alerts issued by the early warning system, and define detailed impact scenarios that allow evacuation plans and mitigation strategies to be developed with local authorities (e.g. civil protection, municipalities). Such strategies require knowledge of the local context and a detailed definition of hazard/risk scenarios compatible with the potential tsunamis that may affect the region under consideration. Recent studies have been conducted for Southern Italy both on expected inundation assessment (Tonini et al., 2021) and tsunami risk perception (Cerase et al., 2019). However, similar research has not yet been carried out for the Northern Adriatic, where the bathymetric and topographic setting is substantially different. The available tsunami hazard estimates (e.g. NEAMTHM18 maps, Basili et al., 2019), in fact, are provided for sites located along the 50 m depth isobaths; given the very shallow bathymetry of the Northern Adriatic, these sites are very far from the areas of interest, at a distance of more than one hundred kilometres from the coastline (e.g., Amato et al. 2021 and references therein), as shown in Figure 2. Usually, in order to calculate the *run-up* to the coastline (i.e. the maximum topographic height, compared to the mean sea level, reached by the tsunami wave during its ingression), a coefficient (i.e. a multiplication factor) is applied to the maximum amplitude of the tsunami wave estimated along the 50 m isobaths (Tonini et al., 2021). However, in the Northern Adriatic the distance between these isobaths and the coastline is vast, and such an empirical relationship may well turn out inadequate. In addition, in the alert messages, a minimum wave amplitude (or a run-up) level is given for the area of interest, but an upper bound is not provided.According to the mentioned SiAM directive (see Attachment 1 of the directive), the accuracy and reliability of inundation data can be significantly improved at local scale (e.g., municipal level and port areas), especially for those areas particularly critical due to the presence of strategic infrastructures, high population density, etc., where it is appropriate to integrate its results with detailed studies through physical-numerical modelling. Accordingly, this study aims to provide a more complete and detailed description of the tsunami hazard, according to a multi-scenario approach, for specific locations along the north-eastern coast of the Adriatic Sea, namely the cities of Trieste, Monfalcone, Grado and Lignano (Fig. 2), which are referred hereinafter as Areas of Interest (AOI).

The urban centres selected as AOI for this study (Fig. 2) represent important economic, historical and cultural centres. The city of Trieste, the largest city in the Friuli Venezia Giulia Region, is of fundamental economic importance due to the presence of ports and other infrastructures, which make it a strategic site for trade and tourism, in addition to its high historical relevance. Monfalcone is an industrial site, with several shipyards and a textile and chemical production centre. The cities of Grado and Lignano are among the leading Italian seaside touristic centres. The city of Lignano, in particular, is a primary touristic site, reaching a population of several hundred thousand visitors during the summer period; the considerable variability of the population throughout the year makes it an area of particular interest for future tsunami risk assessment. Moreover, between Grado and Lignano there is an extensive lagoon, characterised by a peculiar and fragile ecosystem, with a high degree of vulnerability, due to its direct connection with the Adriatic Sea basin.

Modelling the hazard associated with possible tsunamigenic earthquakes occurring in the Adriatic Sea, is therefore carried out with the purpose of contributing to tsunami risk assessment and emergency management for the selected urban areas located along the Northeastern Adriatic coasts. This research builds on previous studies (Peresan and Hassan, 2022) and it aims to revisit and update the existing tsunami hazard models for the Northeastern Adriatic coastal cities of Lignano, Grado, Monfalcone, and Trieste. Recent results based on the computation of a wide set of tsunami scenarios are presented, considering the possible sources as defined in the most updated database of seismogenic sources; parametric tests are performed as well, to account for seismic source variability. Specifically, this study is accomplished accounting for the recently updated DISS database of seismogenic sources in the Adriatic Sea, considering different potential tsunamigenic sources of tectonic origin, located in three distance ranges from the areas of interest (namely at Adriatic-wide, regional and local scales). Selected basin-wide sources in the Mediterranean Sea are considered as well, in order to assess the possible relevance to the AOI of tsunamis occurring outside the Adriatic Sea.

Tsunami modelling is performed by the NAMI DANCE software (Yalciner et al., 2006; Zaytsev et al., 2019), which allows accounting for seismic source properties, variable bathymetry, and non-linear effects in waves generation, propagation and inundation. The software has been successfully tested for recent tsunamis in the Mediterranean area (e.g. Dogan et al., 2021). Three nested-grids of topo-bathymetry are used in the computations, i.e. a coarse grid of about 400 m (Adriatic and Mediterranean scale), a medium grid of about 111 m (North Adriatic), and a fine grid of 20 m (area of interest).Accordingly, a set of tsunami-related parameters and maps are obtained (e.g. maximum tsunami wave amplitude, arrival times, synthetic mareograms), relevant towards planning mitigation actions at the selected sites of the Northeastern Adriatic.

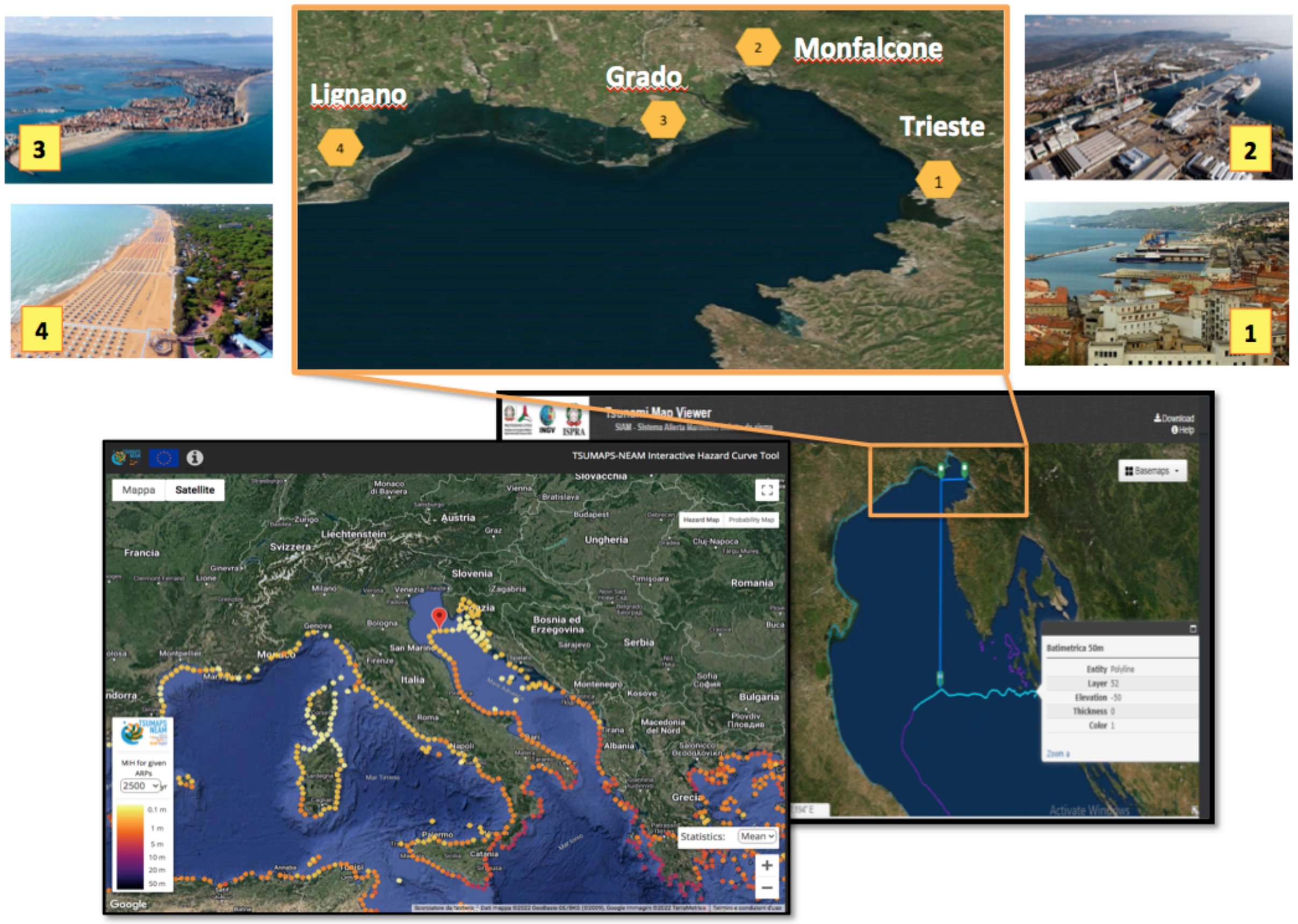

Figure 2 - Regional and zoomed maps of the study region: (top) map of Northeast Adriatic Sea and photos of the areas of interest, AOI; (bottom) map of tsunami hazard estimates for the NEAM macro-area "Northeastern Atlantic, the Mediterranean, and connected seas" (NEAMTHM18; http://ai2lab.org/tsumapsneam/interactive-hazard-curve-tool/, last accessed on March 2022), and map showing the distance between the sites where hazard estimates are provided (50 m isobaths) and the areas of interest.

This work is composed by two main parts. In the first part we evaluate the hazard from the potential tsunamigenic sources located within the Adriatic Sea, which are categorized in this study into Adriatic-wide, regional and local sources, based upon their distance from the site of interest and in view of the tsunami decision matrix currently in force (Fig. 1). We provide maps of maximum tsunami wave amplitude for the whole grid till the shoreline of the area of interest, and near-shore wave amplitude at point of interests (POIs) for local, regional and Adriatic scales. Also, we provide the same maps for an aggregated scenario extracted from all possible scenarios within the Adriatic Sea, at whatever distance. In the second part, we compute the tsunami scenarios for possible very distant sources located outside the Adriatic Sea, hereinafter called basin-wide, from five main tsunamigenic earthquake zones (West Greece, Calabrian, West Hellenic, East Hellenic and Cyprian Arcs). The tsunami scenarios of maximum expected earthquake (based on historical seismicity or geological surveys) within each of these sources have been computed and an aggregated scenario has been extracted. Moreover, maps of maximum wave amplitude for aggregated scenarios at basin and area of interest scales are given for each tsunamigenic earthquake zone.

## 2. Tsunami modelling

Tsunami modelling usually relies on a specific model, namely a mathematical formulation that describes the physical characteristics of the tsunami, to compute the generation and propagation of tsunami waves and their coastal impact. The tsunami modelling process comprises three phases: a) source modelling, in which the initial conditions of tsunami source is computed (i.e., initial sea surface displacements and velocity fields); b) propagation modelling, to be used to describe propagation of tsunamis in the open sea; c)inundation modelling, in which coastal modification of tsunami waves and run-up,in shallow regions and on land, is treated. The last phase is possibly the most important in tsunami modelling, as suitable modelling tools and data are essential for an accurate quantification.

Numerical models (i.e., computer-derived simulation packages) generally use a grid system for the area of interest that includes information about the source, bathymetry, topography and surface roughness or land use. Therefore, a numerical model can incorporate complicated geographic variations in bathymetry, topography and land uses, and can simulate different stages of tsunami, including their variations in wave amplitude, current speed and inundation depth.The recent tsunami models simulate tsunami generation, propagation and inundation stages, overcoming the challenge of abrupt changes in conditions at the shore, which is the most dynamic and complex stage of tsunami modelling process.Among the available models, COMCOT (Cornell University, USA; GNS Science, New Zealand), TSUNAMI-N1/N2 (Tohoku University, Japan) and MOST (National Center for Tsunami Research, USA) have demonstrated their capability for the investigation of the three stages of tsunami evolution. In this study we perform the tsunami numerical modelling by the code NAMI DANCE (Middle East Technical University-METU, Turkey), which represents an extension and update of the model TSUNAMI-N1/N2, and is based on the solution of nonlinear form of the long wave equations with respect to related initial and boundary conditions. In general, the most common and proper solution, upon which the aforementioned codes are developed, is the numerical solution in Nonlinear Shallow Water approximation (Imamura et al., 1988), which neglects the vertical variation of velocity over water depth, due to the fact that tsunami wave lengths are usually much larger than ocean depths.

The tsunami modelling computation carried out in this study is based on non-linear Cartesian shallow water equation. The role of roughness is considered by attributing a friction coefficient of 0.015, taken from Chow (1960).As far as the source model is concerned, according to the manual of NAMI DANCE code (Zaytsev et al., 2019 and references therein), the initial condition for the shallow water wave equation is computed following Okada (1985), which provides the numerical result of the elastic seabed displacement of the co-seismic deformation. For the purpose of source modelling, a rectangular fault is used, which is defined by the following parameters:strike, dip, rake, depth, amount of slip, length L and width W, along with its epicentral coordinates. The Okada (1985) model for fault slip is linear and hence, if necessary, it is possible to model a complex rupture as a group of fault segments, each with a different amount of slip. However, Okada's model neglects the spatiotemporal variations in slip, which are very important for assessing tsunami impacts in regions close to the earthquake source. Geist (1998) showed that local tsunami run-up can vary by over a factor of 3, depending on the slip distribution. Therefore, spatiotemporal variation in fault rupture should be considered when treating the local tsunamigenic sources in the Gulf of Trieste.

## 3. Topo-bathymetry data

Three nested-grids of topo-bathymetry have been used in the computations, namely: a coarse grid of about 400 m spatial resolution, at the Adriatic and Mediterranean scale; a medium grid of about 111 m resolution for North Adriatic; a fine grid of 20 m resolution for the area of interest.

The topo-bathymetry data were extracted, inspected, processed and then converted into the format required by the tsunami modelling code (NAMI DANCE). The coarse-domain bathymetry grids of about 400 m spatial resolution (parent) for the eastern Mediterranean and Adriatic seas are extracted from GEBCO-2020 (GEBCO Compilation Group 2020)). Also, medium-domain of about 111 m resolution grids are obtained and processed from the EMODnet portal (https://portal.emodnet-bathymetry.eu/). Finally, a fine-bathymetry of about 10 m resolution for Northeastern Adriatic, where the municipalities of interest (AOI) are located, is provided by Trobec et al. (2018), who used various geophysical datasets to develop bathymetry model of the Gulf of Trieste (Fig. 3). It is worth mentioning that in this study the grid of 10 m resolution was de-sampled into a 20 m resolution grid, due to current computational limits;the lower resolution grid, however, might not allow to capture some important features of the wetlands within the area of interest(e.g., Marano Lagoon, between Lignano and Grado). A careful examination of the considered bathymetry dataset, in fact, revealed some artifacts (e.g., closed openings) that could affect the propagation of tsunami waves behind the shoreline. For instance, the closed openings to lagoons that appear in the de-sampled bathymetry dataset may prevent tsunamis from progressively propagating within the wetlands. Therefore, the bathymetry data for these areas require careful preliminary analysis, and should be properly modified to closely reflect reality,in order to allow for the computation of tsunami inundation. A focused and detailed mapping is foreseen in these specific areas, the Marano Lagoon in particular, and must be developed to allow for effective exposure, vulnerability and risk analysis at some of the sites of interest.

The topo-bathymetry maps of the study region (Fig. 3) indicate that the deeper part is located in the southern part of the Adriatic Sea, while the northern part is very shallow with water depth less than 50 m. Remarkably, due to the shallow bathymetry the tsunami waves propagate with very low velocity in northern Adriatic Sea. This a major advantage especially in case of distant tsunamigenic sources, as more time is available to alert people and to take emergency actions.

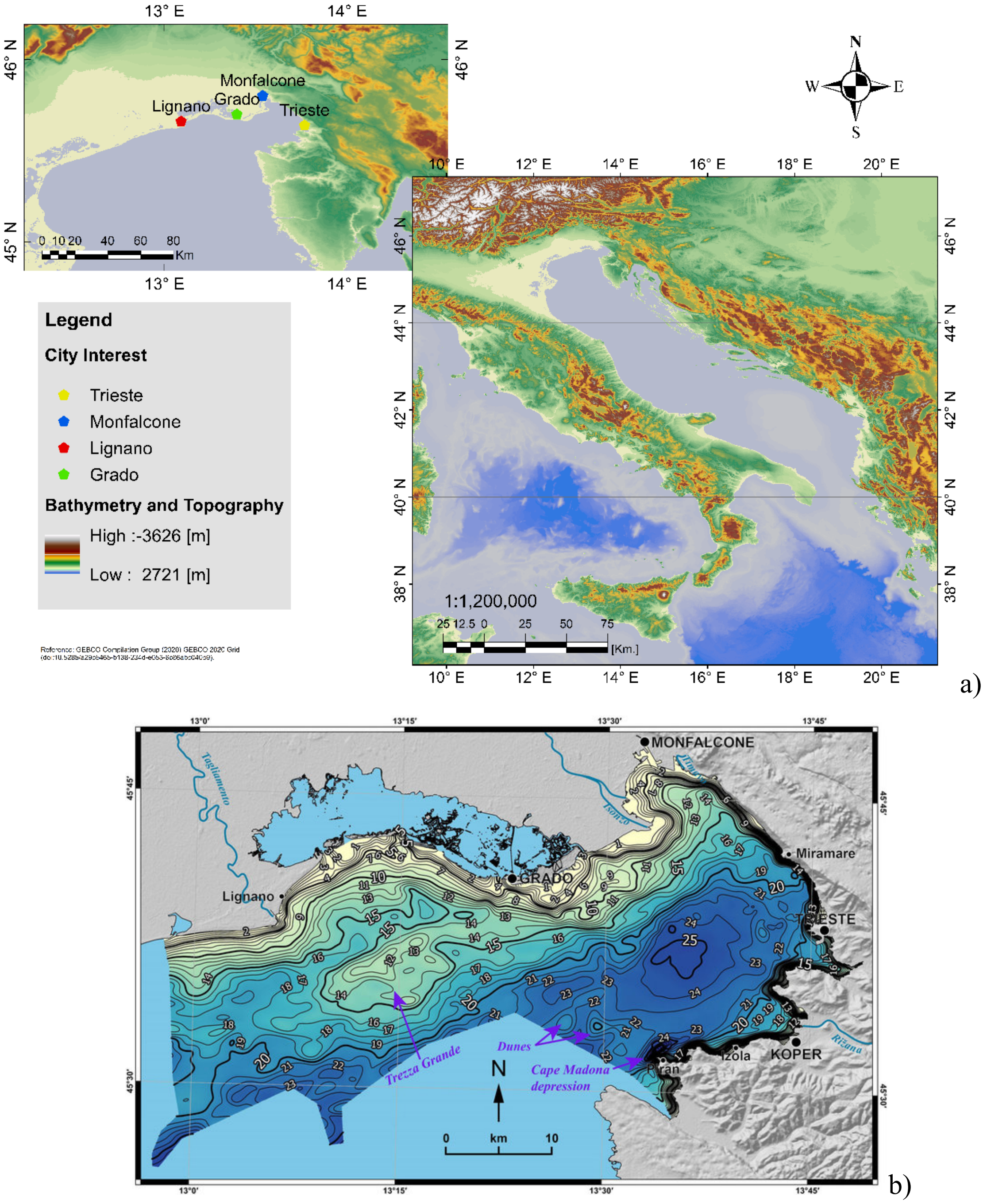

Figure 3 - Bathymetry and topography of the study area and of the area of interest. The large domain, with coarse bathymetry is taken from GEBCO-2020 while the smaller domain is taken from EMODnet (a) and Trobec et al., (2018) (b).

## 4. Computation of tsunamigenic earthquake scenarios in the Adriatic Sea

One of the goals of this study is providing physically consistent tsunami hazard estimates,based on modelling of tsunami waves propagation from a wide set of possible sources, to be considered in case an alert is issued by the CAT-INGV tsunami warning system. Hence,we first defined the possible tsunamigenic earthquake sources located in the Adriatic Sea, based on the recently updated DISS database (DISS-3.3

http://diss.rm.ingv.it/diss/), and following the tsunami decision matrix developed by CAT-INGV (Fig. 1). According to the decision matrix, a seismogenic source is considered tsunamigenic, of potential weak local tsunami (distance ≤ 100 km), if it is capable to produce shallow offshore earthquakes of magnitude M≥6. Therefore, we compiled a database (see Tab. 1 and Fig. 4)of local tsunamigenic earthquake sources of $M_{max}$≥6 relative to our area of interest from the DISS-3.3 database (Fig.4) and $M_{max}$≥6.5 for regional (100 km < distance ≤ 400) and basin-wide (distance > 400 km) tsunamis,to evaluate the potential tsunami hazard at different distances and to serve as pre-computed scenarios in case a future tsunami alert is activated. In this way, in case of tsunamigenic earthquake occurrence, the results for best matched pre-computed scenario (in location and magnitude) can be identified, extracted and used for a quick response. It worth noting that the maximum magnitude of most seismogenic sources within the DISS-3.3 has been augmented in a range 0-1.7 magnitude units (Tab. 1) relative to the previous DISS-3.2.1 version.Therefore, according to the updated DISS-3.3 a larger number of seismogenic sources turn to be capable to generate tsunami, compared to former DISS version used by Peresan and Hassan (2022), because magnitude increased for several sources.

Based on parametric studies, for each scenario the parameters that compose the worse-case scenario in terms of strike, dip, rake and focal depth were extracted from the given range in the DISS database and then adopted (geometry and kinematic parameters). In addition the fault length, width and slip are estimated from source scaling relationship of Wells and Coppersmith (1994), as reported in Table 1. More than a hundred individual scenarios have been simulated during this work, then maps of maximum tsunami wave amplitude at each grid and POIs near-shore are developed for local, regional and Adriatic-wide tsunamis (the computed scenarios and the related maps are available in digital format).

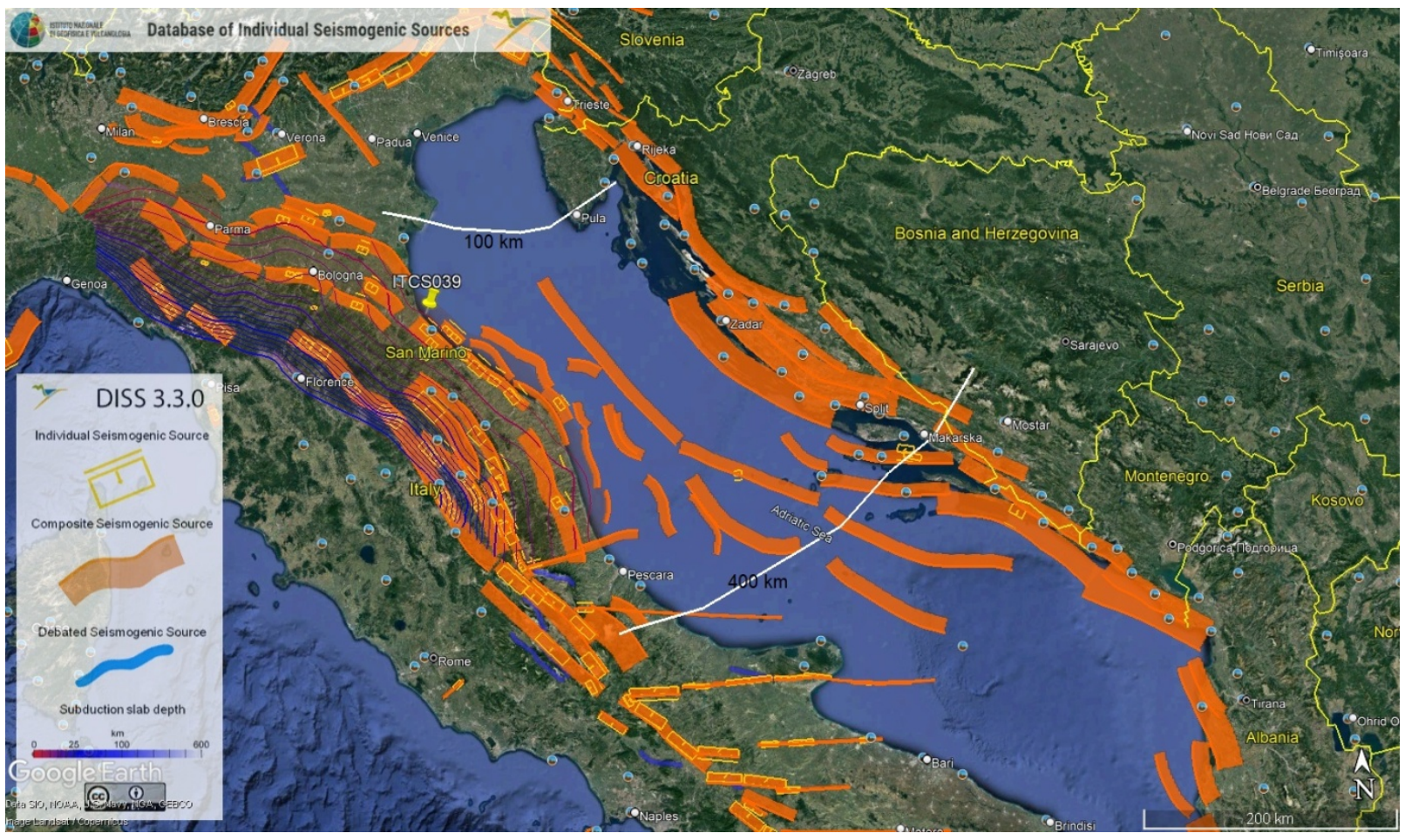

a)

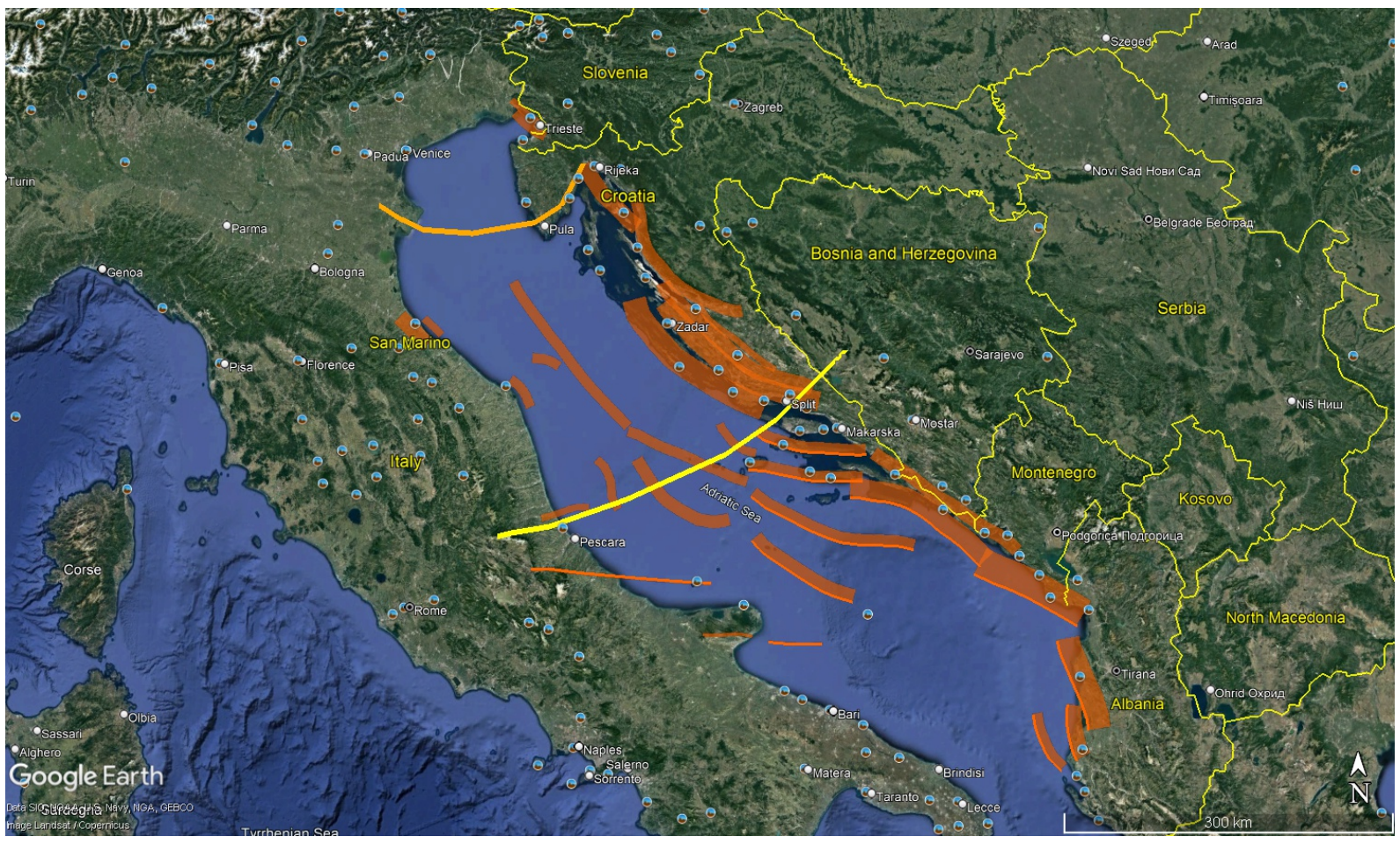

b)

Figure 4 - Seismogenic Sources from DISS-3.3 (DISS Working Group, 2021), including: a) all sources and b) the selected tsunamigenic sources in the Adriatic Sea,located at different location relative the area of interest. Orange and yellow lines indicate the 100 km and 400 km limits respectively.

### 4.1 Adriatic-WideTsunamigenic Earthquake Sources

Following the described procedure, 13 seismogenic sources with $M_{max}$≥6.5, located at distance larger than 400 km from our area of interest,have been selected to compute Adriatic-wide tsunami scenarios. These sources are: ALCS002, ALCS004, ALCS018, ALCS020, HRCS001, HRCS002, HRCS004, HRCS007, HRCS010, ITCS059, ITCS074, MECS001 and MECS007 and their $M_{max}$ is ranging between 6.9 and 8; remarkably, most of the powerful tsunamigenic sources are located along the eastern coast of the Adriatic Sea.The scenarios for the selected sources(Fig. 5) are simulated individually and then an aggregated scenario is developed, as shown in Fig. 6. The computed maximum tsunami wave amplitude at each grid point till shoreline for our AOI is up to 0.5 m, as shown in Fig. 6a.The maximum wave amplitude is provided at given POIs as well, as shown in Fig. 6b, and indicate maximum wave amplitude of 0.5, 0.4, 0.2, and 0.2, for the cities of Lignano, Grado, Monfalcone, and Trieste, respectively. The estimated time of arrivals (ETA) clearly varies from one source to another, with a minimum arrival time in the range 272 - 280 minutes for these cities.

Table 1 - Parameters of offshore tsunamigenic sources in the Adriatic Sea used in tsunami modelling and computation of maps of the maximum tsunami wave amplitude.

| Class | DISS-ID | Name | Updated DISS-3.3 | DISS-3.2 | Difference in $M_{max}$ | Depth [km] | | Strike [deg] | | Dip [deg] | | Rake [deg] | |
|---|---|---|---|---|---|---|---|---|---|---|---|---|---|
| | | | $M_{max}$ | $M_{max}$ | | Min | Max | Min | Max | Min | Max | Min | Max |
| Adriatic-Wide Tsunamigenic Earthquake Sources | ALCS002 | Lushnje | 7.5 | 7.2 | 0.3 | 2.0 | 15.0 | 310.0 | 340.0 | 25.0 | 40.0 | 80.0 | 100.0 |
| | ALCS020 | Seman Coastal | 7.2 | 6.5 | 0.7 | 1.0 | 9.0 | 330.0 | 350.0 | 20.0 | 40.0 | 80.0 | 100.0 |
| | HRCS001 | Mljet | 7.3 | 7.2 | 0.1 | 2.0 | 15.0 | 280.0 | 330.0 | 30.0 | 45.0 | 70.0 | 110.0 |
| | HRCS002 | Hvar | 7.3 | 6.5 | 0.8 | 2.0 | 15.0 | 260.0 | 310.0 | 40.0 | 70.0 | 20.0 | 70.0 |
| | HRCS004 | Eastern Adriatic offshore - South | 6.9 | 6 | 0.9 | 2.0 | 12.0 | 270.0 | 330.0 | 35.0 | 60.0 | 70.0 | 100.0 |
| | HRCS007 | Vis-Korcula | 7.3 | 5.8 | 1.5 | 2.00 | 15.00 | 270 | 290 | 40 | 70 | 20 | 70 |
| | HRCS010 | Palagruza | 6.9 | 6 | 0.9 | 2.0 | 12.0 | 275.0 | 350.0 | 35.0 | 50.0 | 70.0 | 100.0 |
| | ITCS059 | S. Benedetto-Giulianova offshore | 7.5 | 5.8 | 1.7 | 1.00 | 20.00 | 80 | 100 | 65 | 90 | 170 | 230 |
| | ITCS074 | Shallow Gondola Fault Zone | 6.9 | 6.5 | 0.4 | 0.0 | 14.0 | 270.0 | 280.0 | 80.0 | 90.0 | 220.0 | 230.0 |
| | MECS001 | Montenegro offshore | 8 | 7.2 | 0.8 | 2.0 | 15.0 | 290.0 | 310.0 | 15.0 | 40.0 | 80.0 | 100.0 |
| | MECS007 | Budva Offshore | 7.1 | 6.5 | 0.6 | 1.0 | 10.0 | 270.0 | 350.0 | 25.0 | 35.0 | 80.0 | 110.0 |
| Regional tsunamigenic earthquake sources | ITCS031 | Conero offshore | 6.6 | 6 | 0.6 | 1.5 | 7.0 | 125.0 | 155.0 | 25.0 | 40.0 | 80.0 | 100.0 |
| | HRCS021 | Eastern Adriatic offshore - North | 6.9 | 6 | 0.9 | 2.0 | 12.0 | 270.0 | 330.0 | 35.0 | 60.0 | 70.0 | 100.0 |
| | HRCS014 | Jana-1 | 6.9 | 6 | 0.9 | 2.0 | 12.0 | 275.0 | 350.0 | 35.0 | 50.0 | 70.0 | 100.0 |
| | HRCS020 | Eastern Adriatic offshore - Central | 6.9 | 6 | 0.9 | 2.0 | 12.0 | 270.0 | 330.0 | 35.0 | 60.0 | 70.0 | 100.0 |
| | HRCS008 | Dugi Otok | 7.5 | 6 | 1.5 | 2.0 | 18.0 | 280.0 | 330.0 | 30.0 | 45.0 | 90.0 | 120.0 |
| | HRCS018 | Vis-West | 6.9 | 5.8 | 1.1 | 2.00 | 15.00 | 280 | 330 | 50 | 70 | 90 | 120 |
| | HRCS025 | Krk | 7.2 | 6 | 1.2 | 1.0 | 15.0 | 320.0 | 340.0 | 40.0 | 50.0 | 110.0 | 130.0 |
| | ITCS004 | Castelluccio dei Sauri-Trani | 7.3 | 6.3 | 1 | 11.00 | 22.50 | 260 | 280 | 70 | 90 | 170 | 190 |
| | ITCS039 | Riminese onshore | 7 | 5.9 | 1.1 | 2.00 | 10.00 | 120 | 140 | 25 | 35 | 80 | 100 |
| | ITCS070 | Deep Gondola Fault Zone | 7.2 | 6.5 | 0.7 | 0.0 | 14.0 | 270.0 | 280.0 | 80.0 | 90.0 | 220.0 | 230.0 |
| | ITCS106 | Pesaro mare-Cornelia | 6.5 | 5.5 | 1 | 2.00 | 7.00 | 120 | 160 | 25 | 40 | 80 | 100 |
| | ITCS108 | Clara | 6.5 | 5.5 | 1 | 2.00 | 8.00 | 95 | 120 | 30 | 40 | 80 | 100 |
| | ITCS134 | **Roseto degli Abruzzi** | 6.9 | 5.5 | 1.4 | 4.50 | 15.00 | 65 | 75 | 50 | 70 | 120 | 150 |
| | ITCS154 | Edmond | 6.8 | 5.5 | 1.3 | 2.60 | 9.00 | 109 | 190 | 20 | 60 | 80 | 100 |
| | ITCS155 | Daniel | 6.5 | 5.5 | 1 | 3.20 | 10.00 | 299 | 354 | 32 | 75 | 80 | 100 |
| | ITCS158 | Valeria | 6.6 | 5.5 | 1.1 | 4.30 | 10.00 | 294 | 353 | 24 | 48 | 80 | 100 |
| Local tsunamigenic earthquake sources | ITCS100 | Northern Trieste Gulf | 6.9 | 6.5 | 0.4 | 1 | 10 | 320 | 350 | 50 | 60 | 130 | 160 |
| | ITCS101 | Southern Trieste Gulf | 6.5 | 6.5 | 0 | 1.5 | 8 | 290 | 330 | 30 | 45 | 100 | 120 |

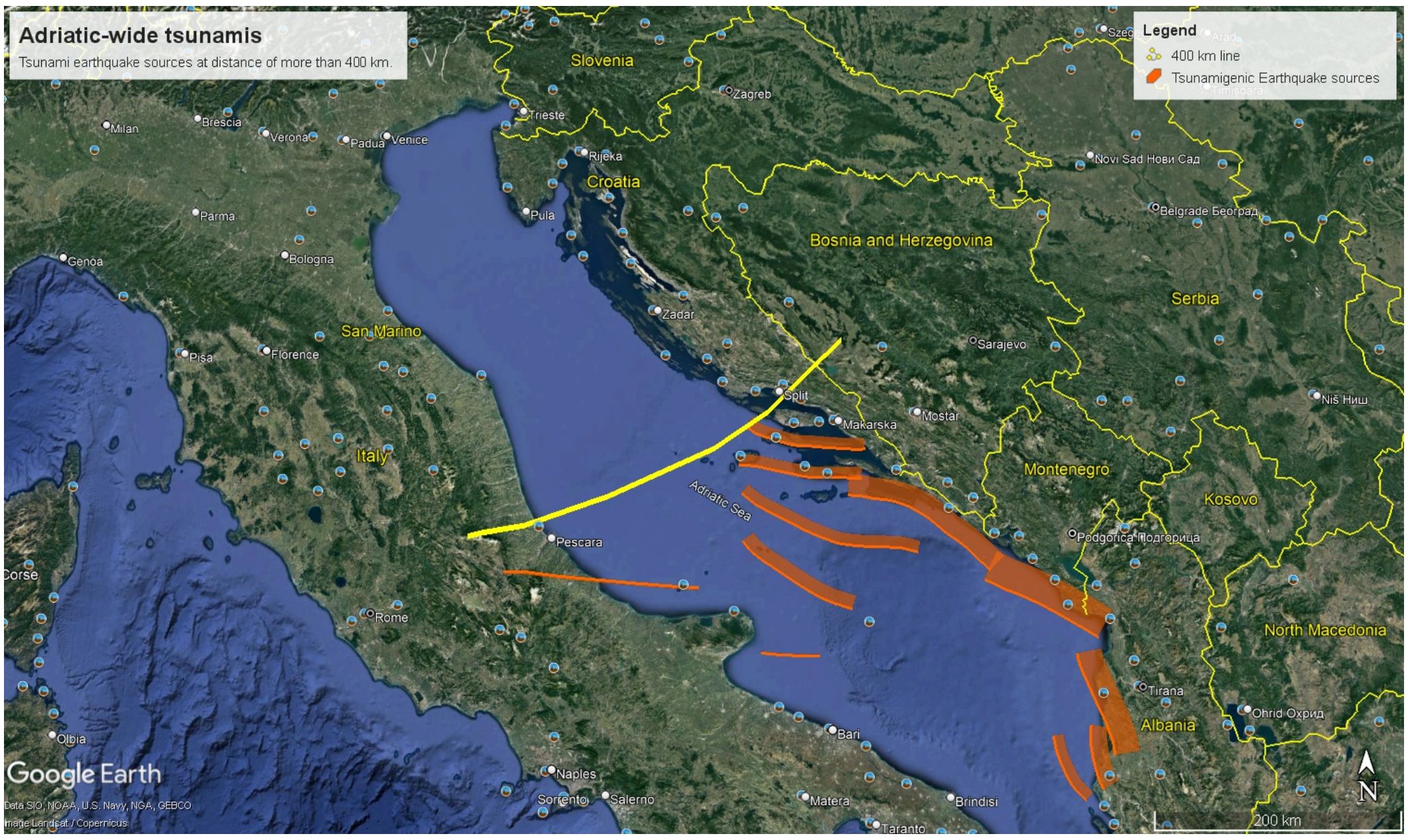

Figure 5 - Tsunamigenic earthquake sources at Adriatic-wide scale (distance from the AOI is >400 km). Yellow line defines the sources located at more than 400 km while red rectangles are the tsunamigenic sources.

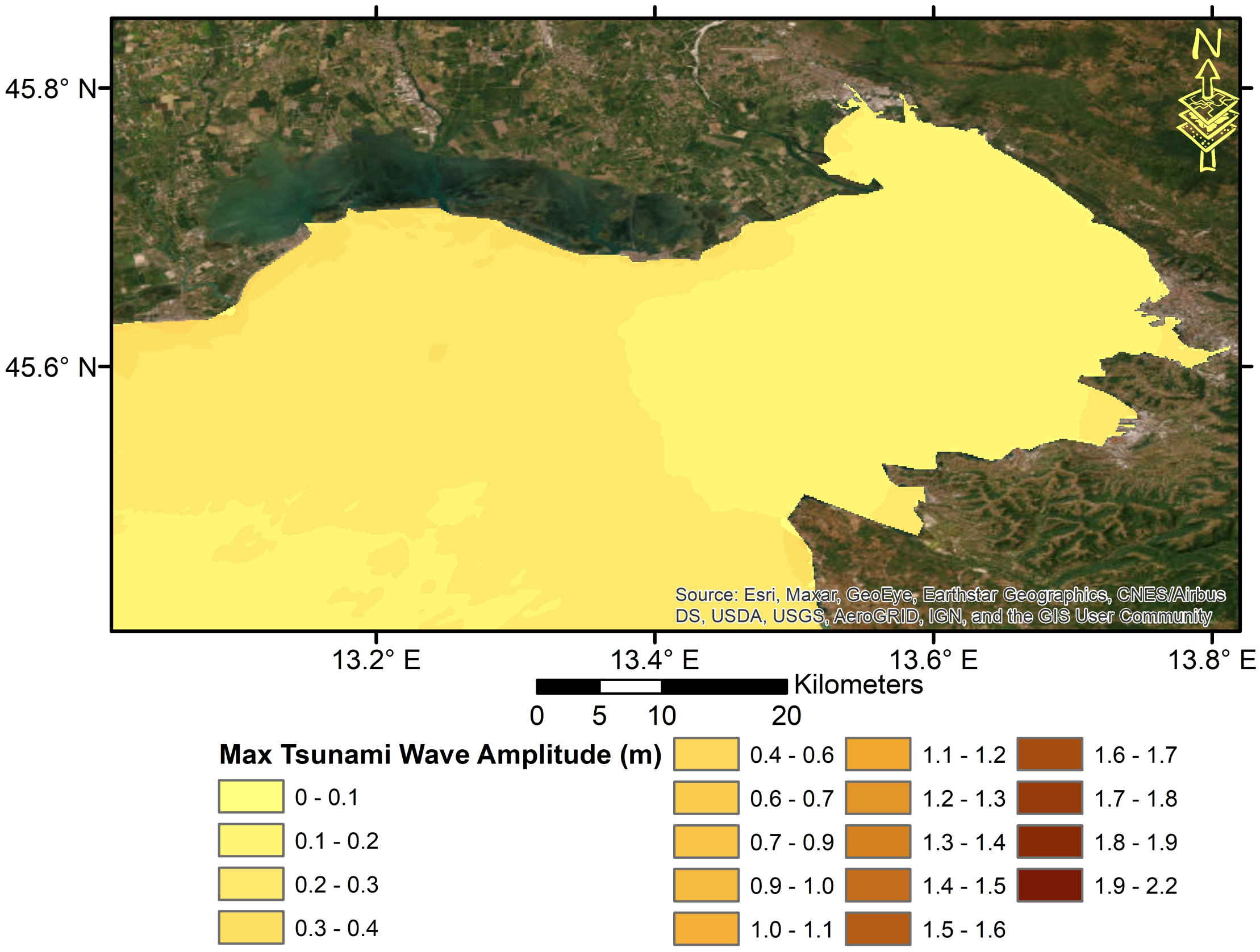

Figure 6a - Map of the maximum tsunami wave amplitude computed for the Adriatic-wide sources located at distance larger than 400 km from the area of interest (AOI).

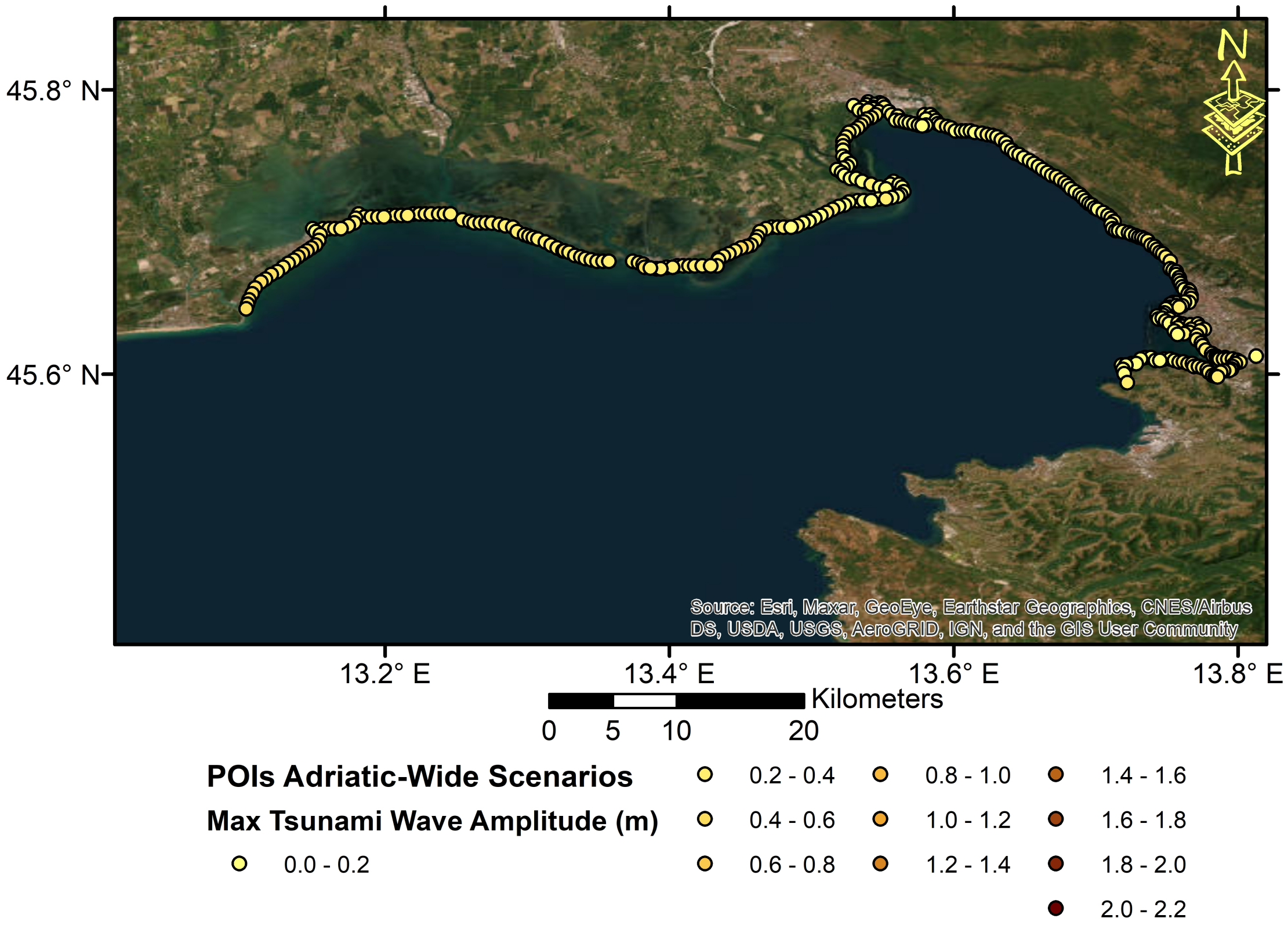

Figure 6b - The maximum tsunami wave amplitude computed for the Adriatic-wide sources at the selected POIs.

### 4.2 Regional tsunamigenic earthquake sources

Hereinafter, following the decision matrix, the seismogenic sources with $M_{max} \geq 6.5$ located at distance between 100 and 400 km from our area of interest, have been considered. Accordingly, more than 15 seismogenic sources from DISS have been defined as potentially tsunamigenic (Fig. 7); they have been simulated individually and then an aggregated scenario has been developed (Fig.8). The computed maximum tsunami wave amplitude for our AOI is up to 1.0 m at some sites,as shown in Fig. 8a. The regional tsunami sources are found to impose a higher hazard to the AOI, compared to the Adriatic-wide tsunami sources. The maximum wave amplitude along the shoreline is provided at given POIs as well, as shown in Fig. 8b.The simulated scenarios indicate a maximum wave amplitude of 0.9, 1.0, 0.4, and 0.4 meters, for the cities of Lignano, Grado, Monfalcone, and Trieste, respectively. Moreover, the minimum arrival time ETA is ranging between 107 to 115 minutes for these cities.

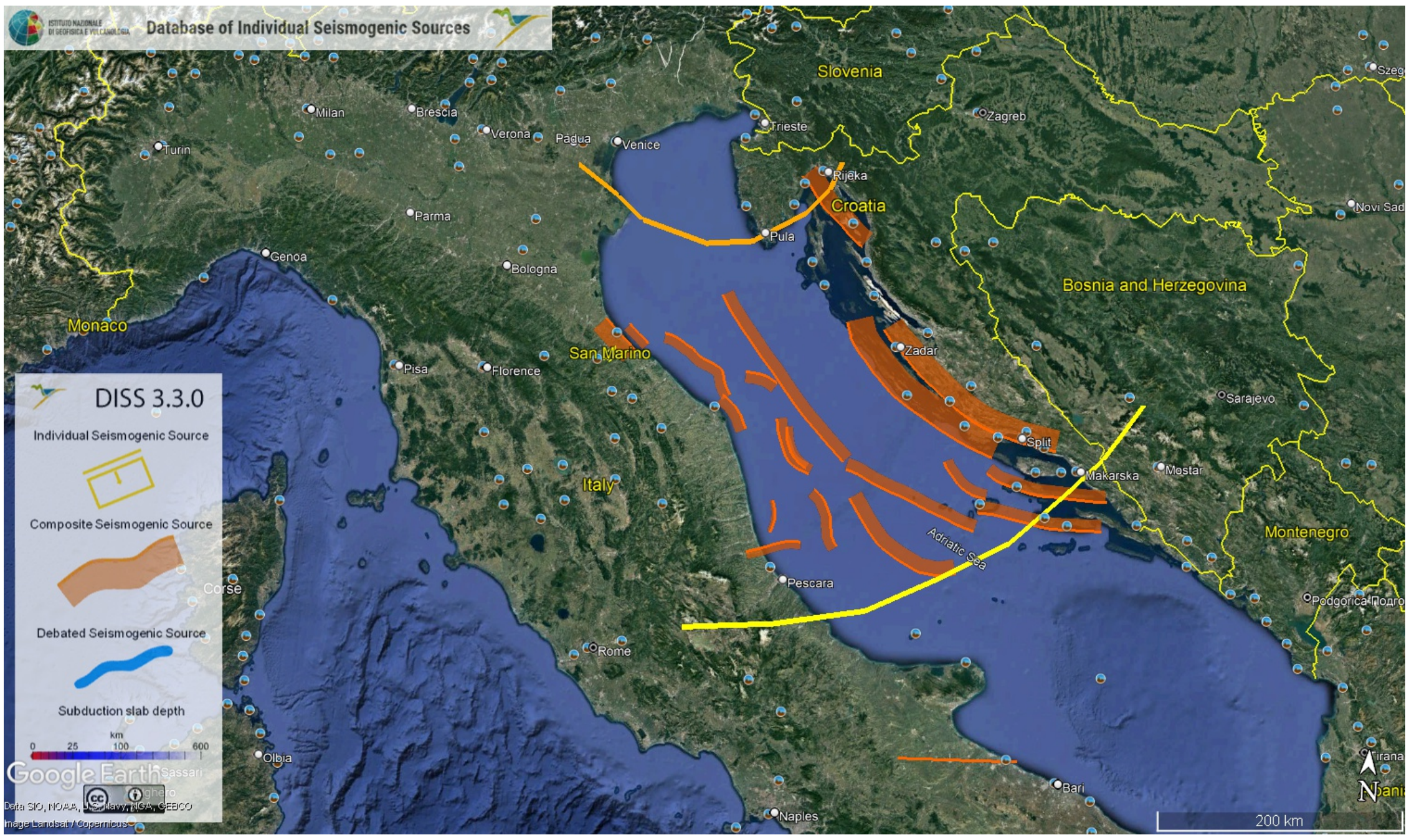

Figure 7 - Tsunamigenic earthquake sources at Regional scale (distance from the AOI is >100 and ≤400 km). Orange line defines 100 km epicentral distance while Yellow line defines the 400 km epicentral distance. Also, while red rectangles represent tsunamigenic sources.

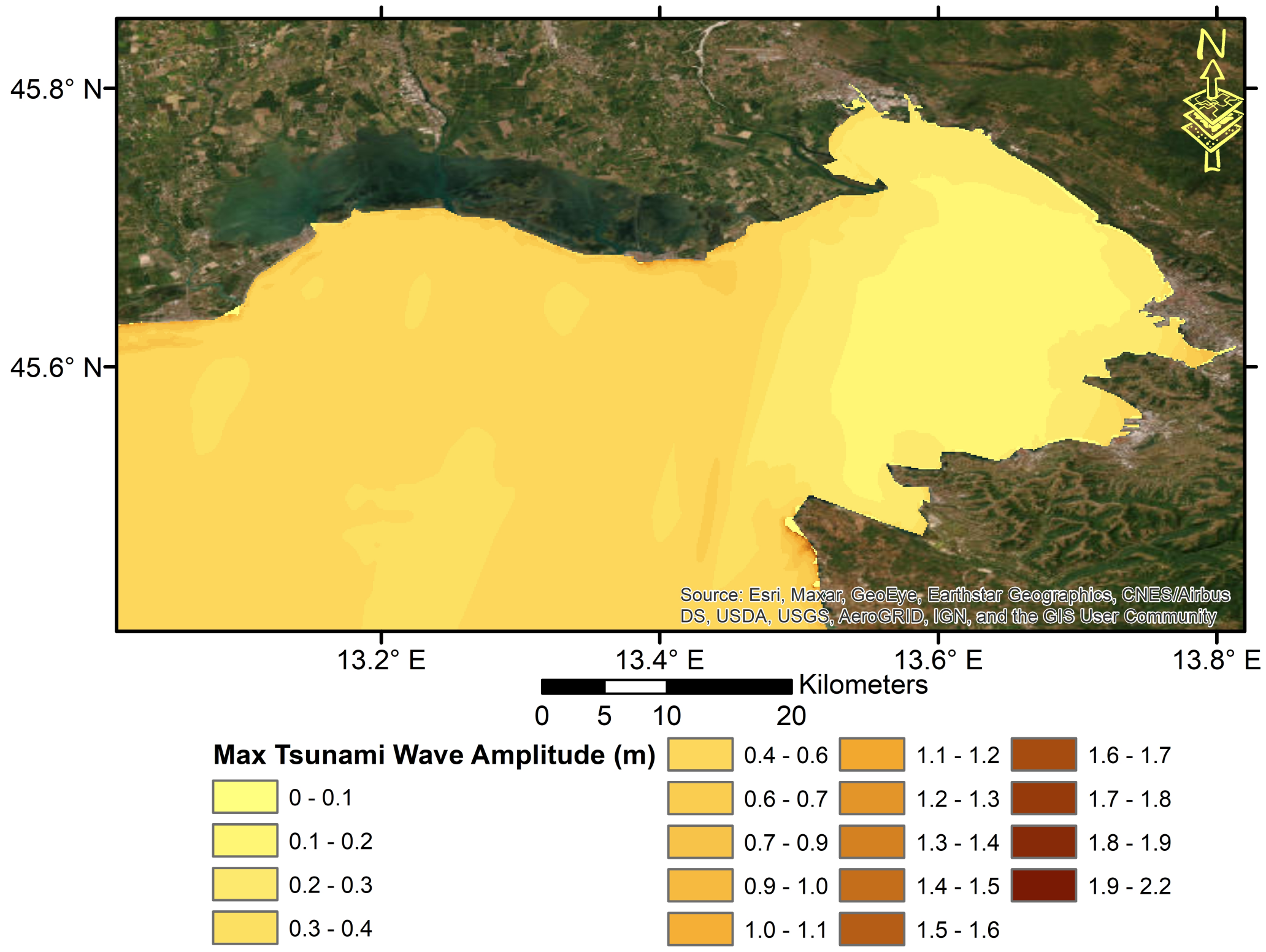

Figure 8a - Map of the maximum tsunami wave amplitude computed for the regional tsunamigenic sources.

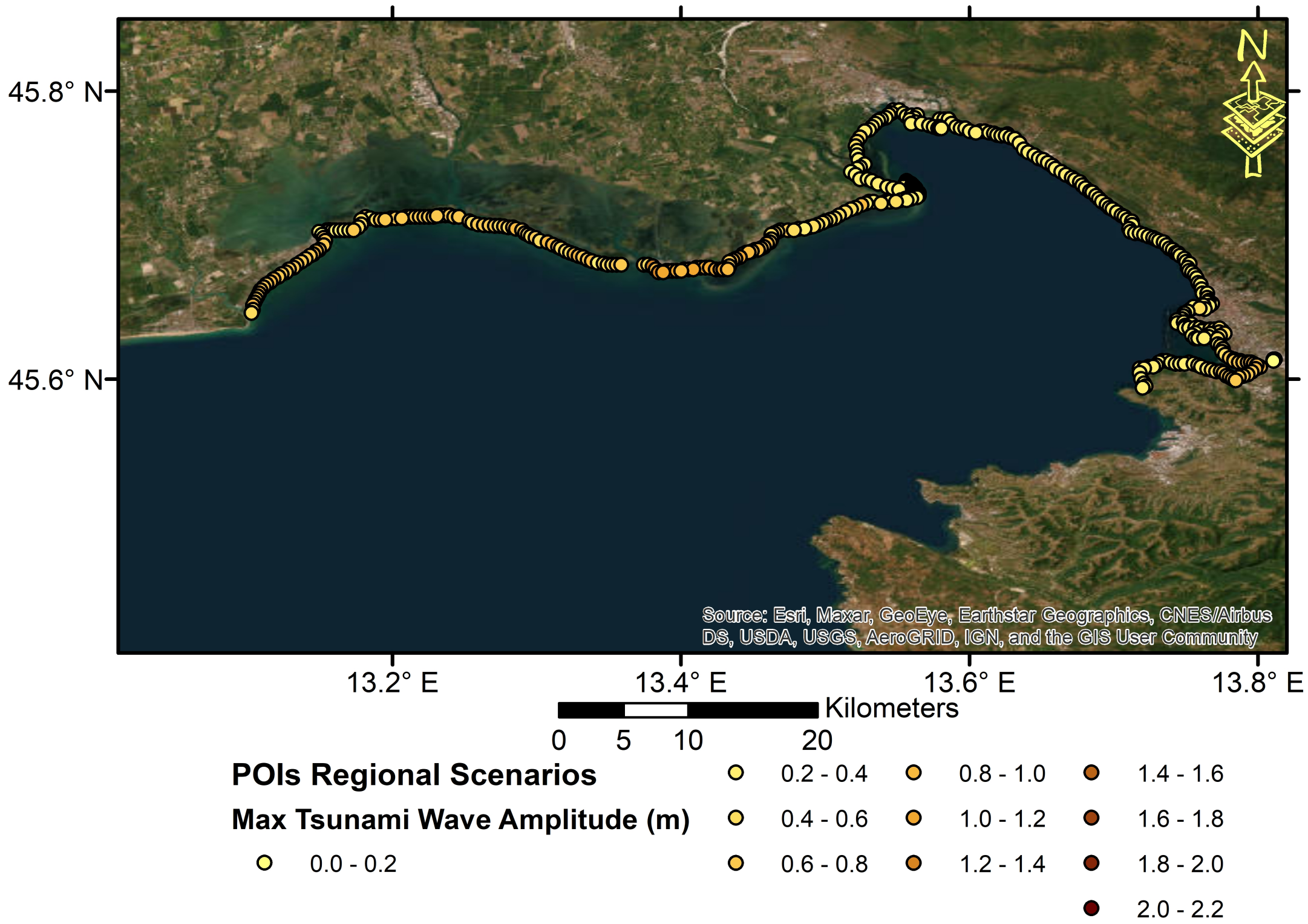

Figure 8b - The maximum tsunami wave amplitude computed for the regional tsunamigenic sources at POIs.

### 4.3 Local tsunamigenic earthquake sources

Two local offshore seismogenic sources are defined in the north-east Adriatic region according to DISS-3.3 database(Tab. 1 and Fig.4; for further details, see DISS Working Group 2021).These earthquake sources are named the Southern and the Northern Trieste Gulf, as indicated in Table 1, and have assigned magnitudes $M_{max}$ 6.5 and 6.9, respectively. These two sources are considered tsunamigenic because of their assigned location, magnitude and focal mechanism, and they might impact the AOI with strong ground motion and tsunami wave if they occur.

In the current study, the local seismogenic sources are considered in developing the tsunami hazard estimation. The offshore source is segmented into sub-sources of 7 km length and different slip and rupture velocity increasing in unilateral direction NW. This preliminary source shall be studied in detail in the future developments, considering different rupture styles (i.e. unilateral, bilateral) and directivity angles, which represent different potential scenarios that may occurs in future rupture.The computed maximum tsunami wave amplitude at POIs is up to 2.0 m, as shown in Fig. 9. Specifically, a maximum wave amplitude of 2.0 m is estimated for Trieste and Grado, while Monfalcone and Lignano have 1.0 and 0.4 m, respectively.

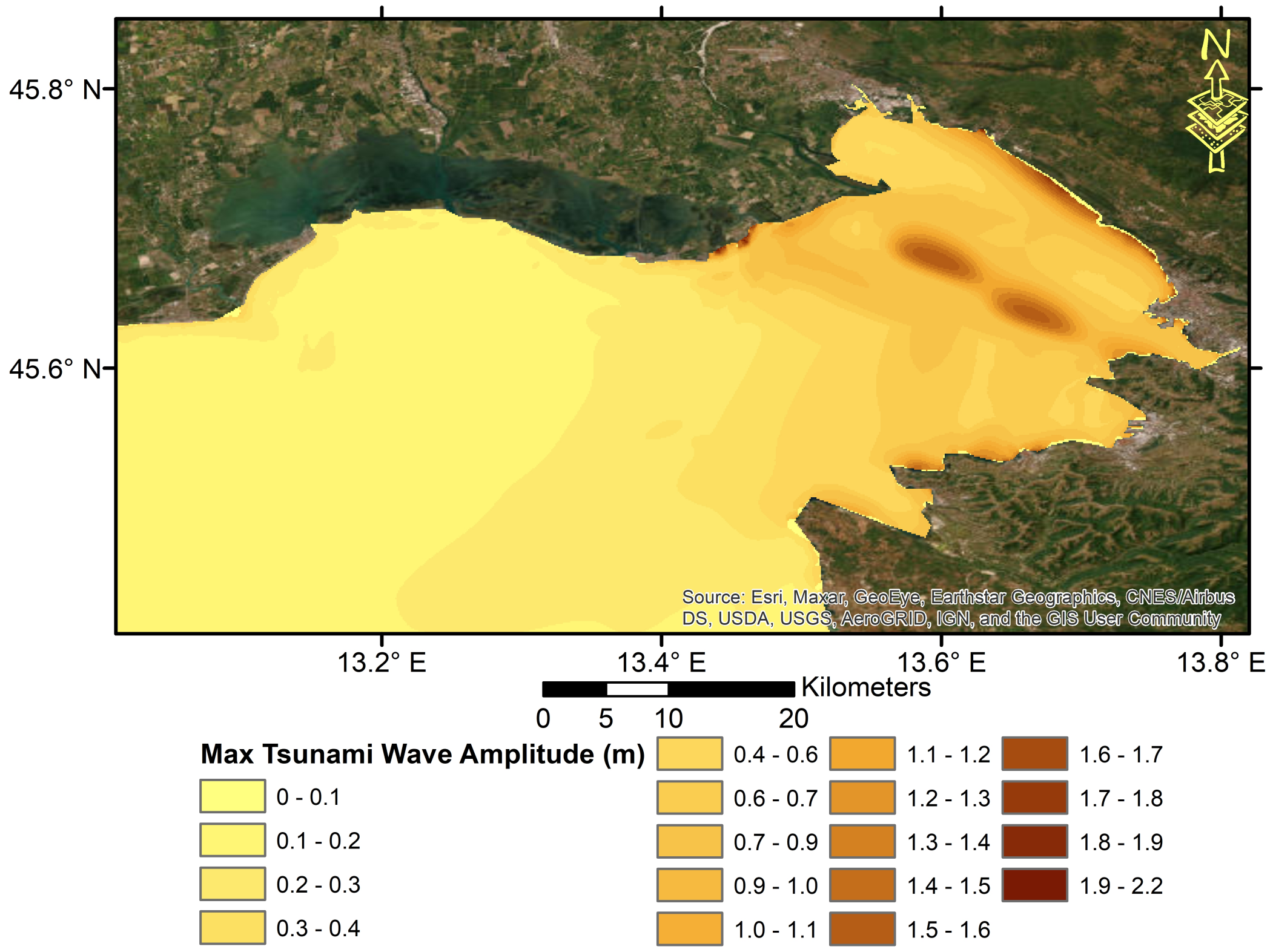

Figure 9a - The maximum tsunami wave amplitudes for local sources located at distance less than 100 km from the area of interest.

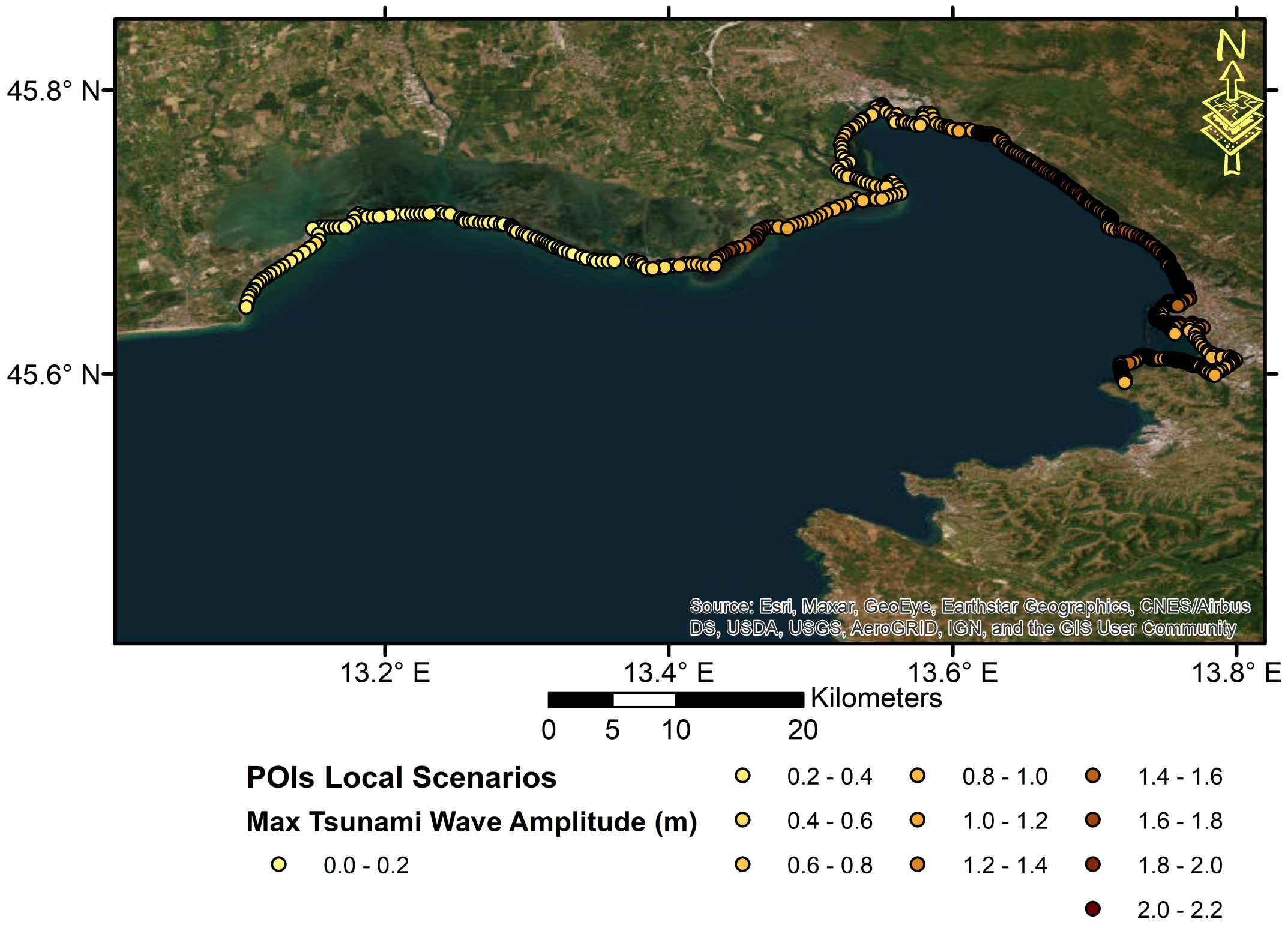

Figure 9b - The maximum tsunami wave amplitude computed for the local tsunamigenic sources at POIs.

5. **Basin-Wide tsunami sources (out of the Adriatic Sea)**

Based on the CAT-INGV's tsunami matrix (Fig. 1), an offshore or inland, shallow earthquake (i.e. depth <100 km) of magnitude larger than 7.5 has a potential to generate a destructive tsunami affecting the whole Mediterranean basin at any distance. Accordingly, tsunami scenarios have been modelled for a set of potential seismogenic zones of magnitude larger than 7.5 located outside the Adriatic Sea, namely: West Greece, West Hellenic, Calabrian Arc, East Hellenic, and Cyprian Arc zones. GEBCO 2020 topo-bathymetry of about 400 m resolution has been used for the entire basin (Fig. 10). The maximum wave amplitude has been computed for the area of interest in Northern Adriatic (Fig. 11), considering the seismic source parameters provided in Table 2. Fault slip parameters for tsunami scenarios computation were collected both from DISS-3.3 and from individual studies (England et al., 2015; Hassan et al., 2023). For each seismogenic zone separately, a number of scenarios accounting for source uncertainties have been computed and tsunami wave amplitude maps have been developed, as aggregated scenario of maximum estimated values.

The computed scenarios can be used, jointly with real time seismic and sea-level tide measurements, to provide quantitative basis for the decision making process by emergency manger and first responders in FVG, in case of future large tsunamigenic earthquake occurrence in one of the considered zones.

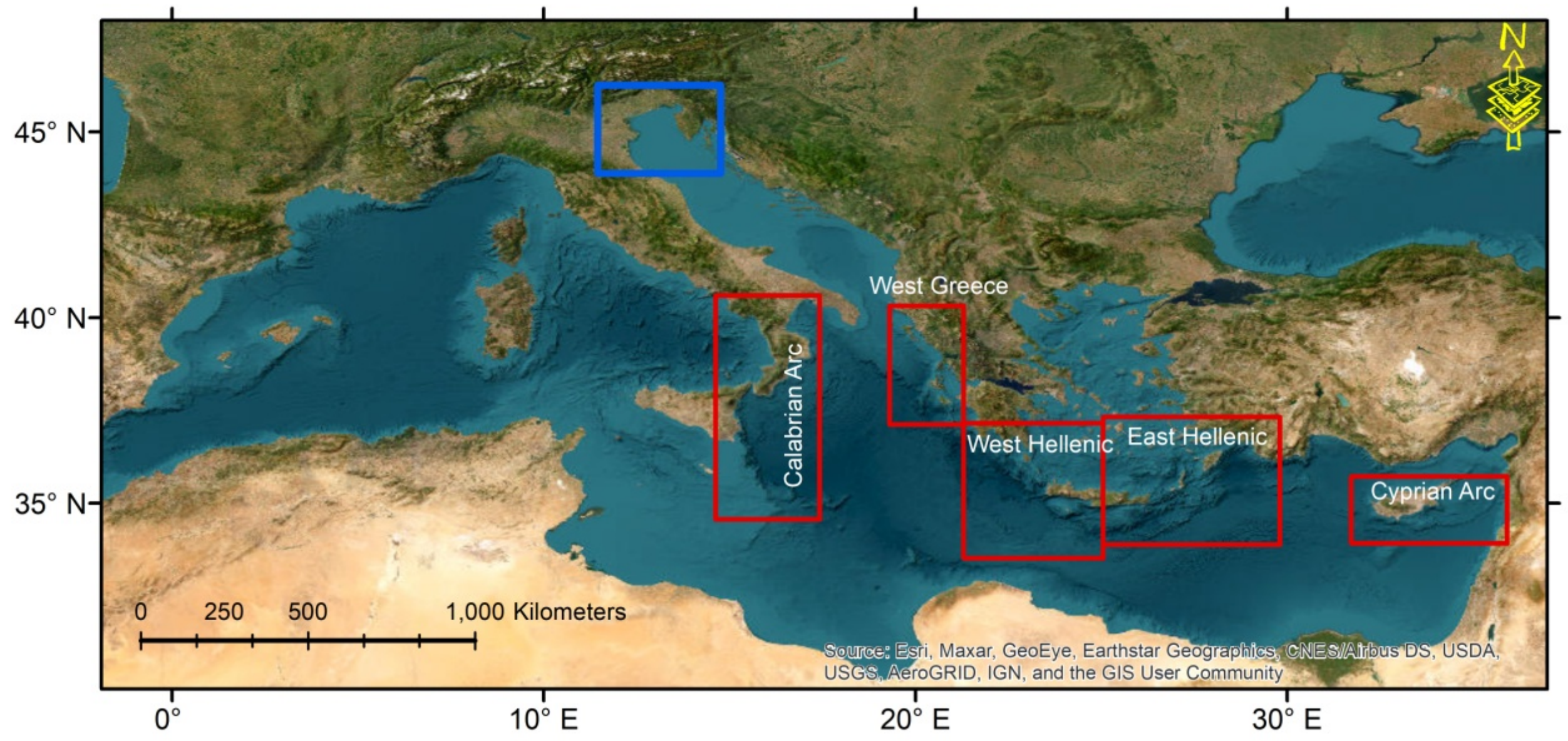

Figure 10 - Basin-wide tsunami sources (red rectangles) within which individual scenarios are modelled and the maps of maximum wave amplitude are computed and presented for the AOI (blue rectangle).

Table 2- Parameters of Basin-Wide tsunamigenic sources in the Mediterranean Sea adopted in the computation of maps of the maximum tsunami wave amplitude.

| Zone | ID | CSS-Name | Mmax DISS-3.3 | Depth [km] | | Strike [deg] | | Dip [deg] | | Rake [deg] | | Reference |
|---|---|---|---|---|---|---|---|---|---|---|---|---|
| West Greece Zone | ITCS096 | Calabria Offshore S | 7.9 | 4 | 12 | 190 | 270 | 10 | 20 | 80 | 100 | DISS-3.3 |
| | ITCS097 | Calabria Offshore SW | 7.9 | 4 | 12 | 180 | 280 | 10 | 20 | 80 | 100 | |
| | ITCS098 | Calabria Offshore NE | 7.4 | 4 | 9 | 180 | 260 | 10 | 20 | 80 | 100 | |
| | MTCS001 | Malta N | 7.5 | 3 | 15 | 120 | 140 | 50 | 75 | 190 | 240 | |
| | ITCS099 | Calabria Offshore NW | 7.9 | 3 | 12 | 180 | 250 | 10 | 40 | 80 | 100 | |
| Calabrian Arc Zone | ALCS001 | Sazani | 7.6 | 1 | 15 | 290 | 340 | 25 | 40 | 80 | 100 | |
| | GRCS606 | Mediterranean North | 7.8 | 5 | 16 | 300 | 360 | 15 | 35 | 80 | 100 | |
| | GRCS607 | Zakinthos Offshore | 7.8 | 5 | 17 | 305 | 360 | 15 | 40 | 80 | 100 | |
| | GRCS604 | Kerkyra Offshore | 8 | 3 | 20 | 280 | 350 | 20 | 40 | 80 | 100 | |
| | GRCS603 | Kephallonia | 7.4 | 4 | 25 | 10 | 50 | 50 | 70 | 140 | 180 | |
| | GRIS001 | Southwestern Crete | 8.5 | 5 | 54 | 300 | 314 | 20 | 35 | 90 | 100 | |
| West Hellenic Arc Zone | S1 | West Hellenic | 8.5 | | 45 | | 326 | | 30 | | 90 | England et al., (2015) |
| | S2 | West Hellenic | 8.5 | | 45 | | 326 | | 35 | | 90 | |
| | S3 | West Hellenic | 8.5 | | 45 | | 326 | | 30 | | 90 | |
| | S4 | West Hellenic | 8.5 | | 45 | | 326 | | 30 | | 90 | |
| | S5 | West Hellenic | 8 | | 45 | | 250 | | 30 | | 90 | |
| East Hellenic Arc Zone | S6 | East Hellenic | 8 | | 45 | | 250 | | 30 | | 90 | |
| | S7 | East Hellenic | 8 | | 45 | | 250 | | 30 | | 90 | |
| | S8 | East Hellenic | 8 | | 40 | | 250 | | 35 | | 90 | |
| Cyprian Arc Zone | S9 | Cyprian Arc | 8 | | 5 | | 280 | | 20 | | 90 | |

## 5.1 West Greece Zone

The western coast of Greece is situated along the Ionian Sea, which is relatively less prone to tsunamis compared to other areas of the Mediterranean Sea, such as the eastern coast of the Mediterranean.However, it is important to note that tsunamis can still occur in the Ionian Sea due to various factors such as earthquakes, underwater landslides, and volcanic eruptions. The most significant event to hit the western coast of Greece in recent history was associated to the 1953 Ionian earthquake, which triggered a tsunami that caused significant damage and loss of life in the region (Mavroulis and Lekkas, 2021).

Five individual tsunami scenarios have been modelled and an aggregated scenario of maximum wave amplitude has been extracted (Fig.11). The resulting maximum wave amplitude at the AOI ranges between 0.10-0.44 m as shown in bottom panel of Fig.11.

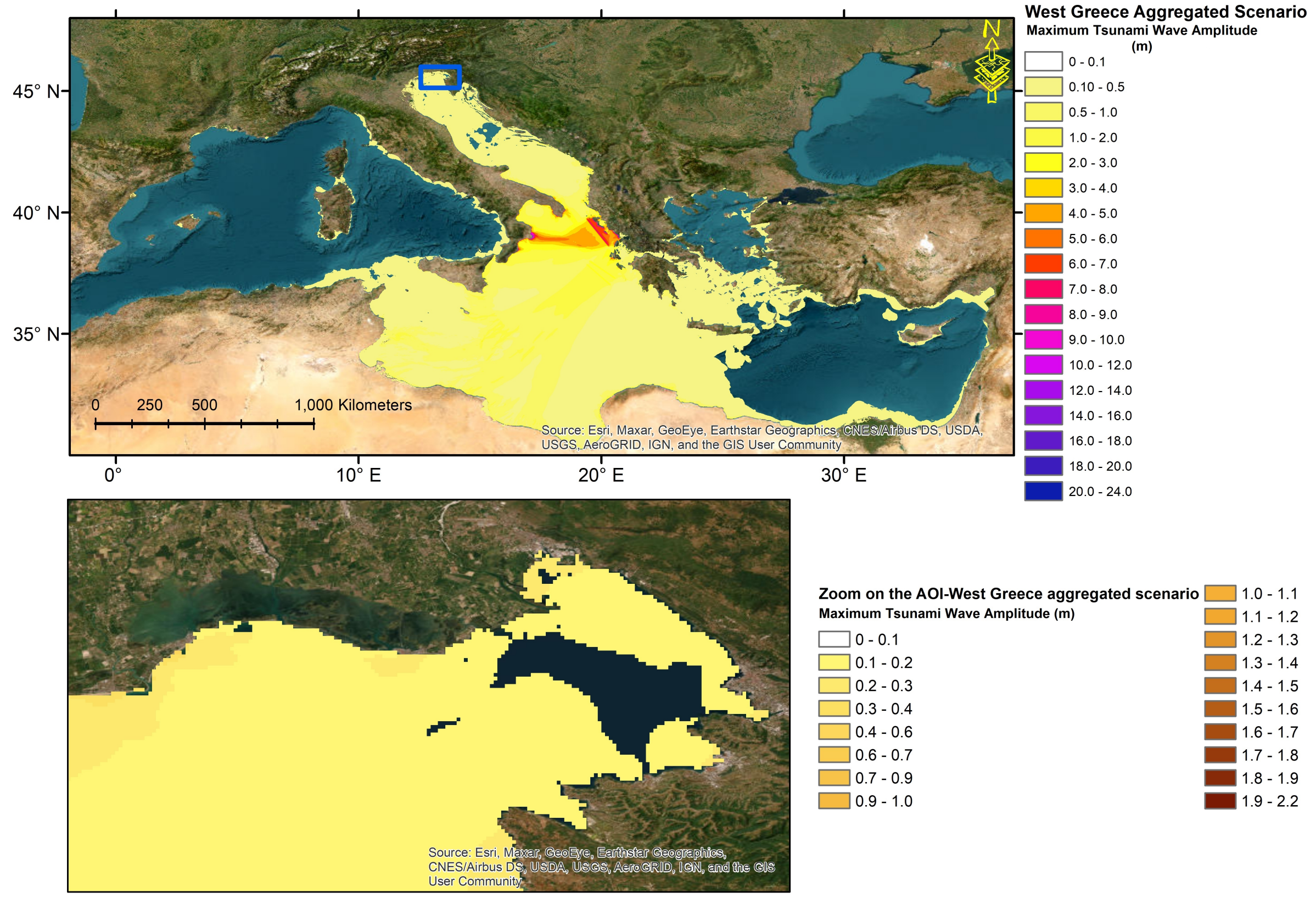

Figure 11 - Aggregated scenario from the individual scenarios computed for the West Greece zone at basin wide scale (top) and a zoom to the area of interest (bottom).

## 5.2 Calabrian Arc Zone

The Calabrian Arc is considered a high-risk area for tsunamis due to seismic activity and underwater landslides. The area experienced several significant tsunamis in the past, including the 1908 Messina earthquake and tsunami (e.g. Schambach et al, 2020), which caused widespread devastation and loss of life in the region.

Eight individual tsunami scenarios have been modelled within the Calabrian Arc zone (Fig. 10), then an aggregated scenario is produced (Fig. 12). The computed maximum near-shore wave amplitude, computed at the AOI, is about 0.2 m as shown in (Fig. 12, bottom panel).

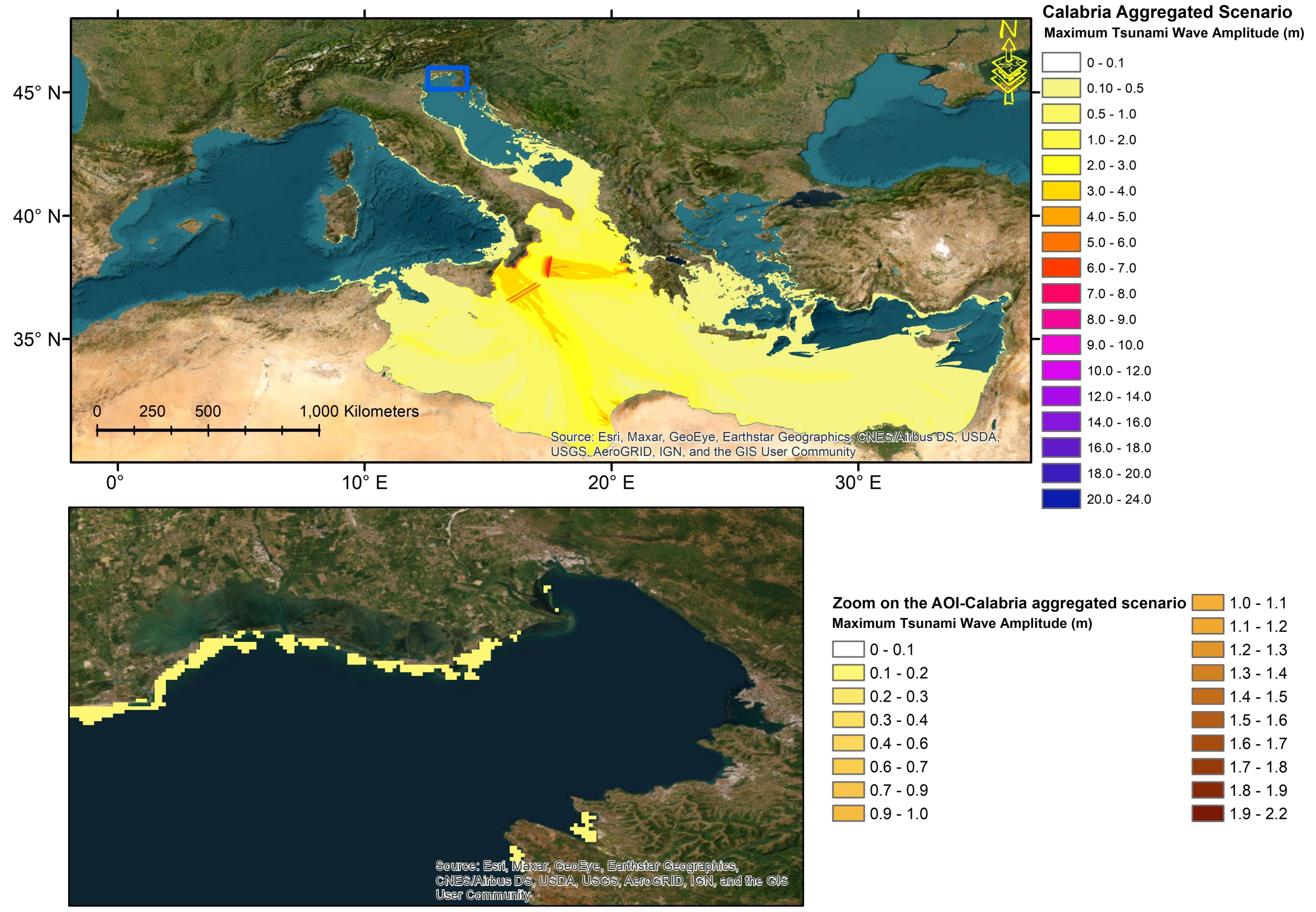

Figure 12 - Aggregated scenario from the individual scenarios computed for the Calabrian Arc zone at basin wide scale (top) and a zoom to the area of interest (bottom).

### 5.3 West Hellenic Arc Zone

The West Hellenic Arc is a seismically active area located in the eastern Mediterranean Sea (Fig. 10). This arc is part of the larger Hellenic Arc, which is a curving chain of tectonic plate boundaries that extends from the western coast of Greece to the southern coast of Turkey. The Hellenic Arc results from the subduction of the African Plate beneath the Eurasian Plate, and is associated with frequent seismic and volcanic activity.The most famous tsunami event was associated with the Mw8.5 Crete earthquake, an extremely powerful earthquake that occurred on July 21, 365 CE, near the island of Crete in the eastern Mediterranean. This earthquake and the related tsunami were among the most devastating events in the past, causing widespread destruction across the Mediterranean region. The tsunami had a far-reaching impact, affecting coastal areas in Africa, the Adriatic, Greece and Sicily (Shaw et al., 2008).

Six individual tsunami scenarios from DISS-3.3 and previous work (Hassan et al., 2023) have been modelled, in order to evaluate the impact on AOI of possible tsunamis originated within this zone, which is capable of generating basin-wide tsunamis (Fig.13). The computed near-shore maximum tsunami wave amplitude at AOI is quite large, and is ranging between 0.15-0.5 m, which makes this source the most powerful compared to the other sources outside the Adriatic Sea(Fig. 13, bottom panel).

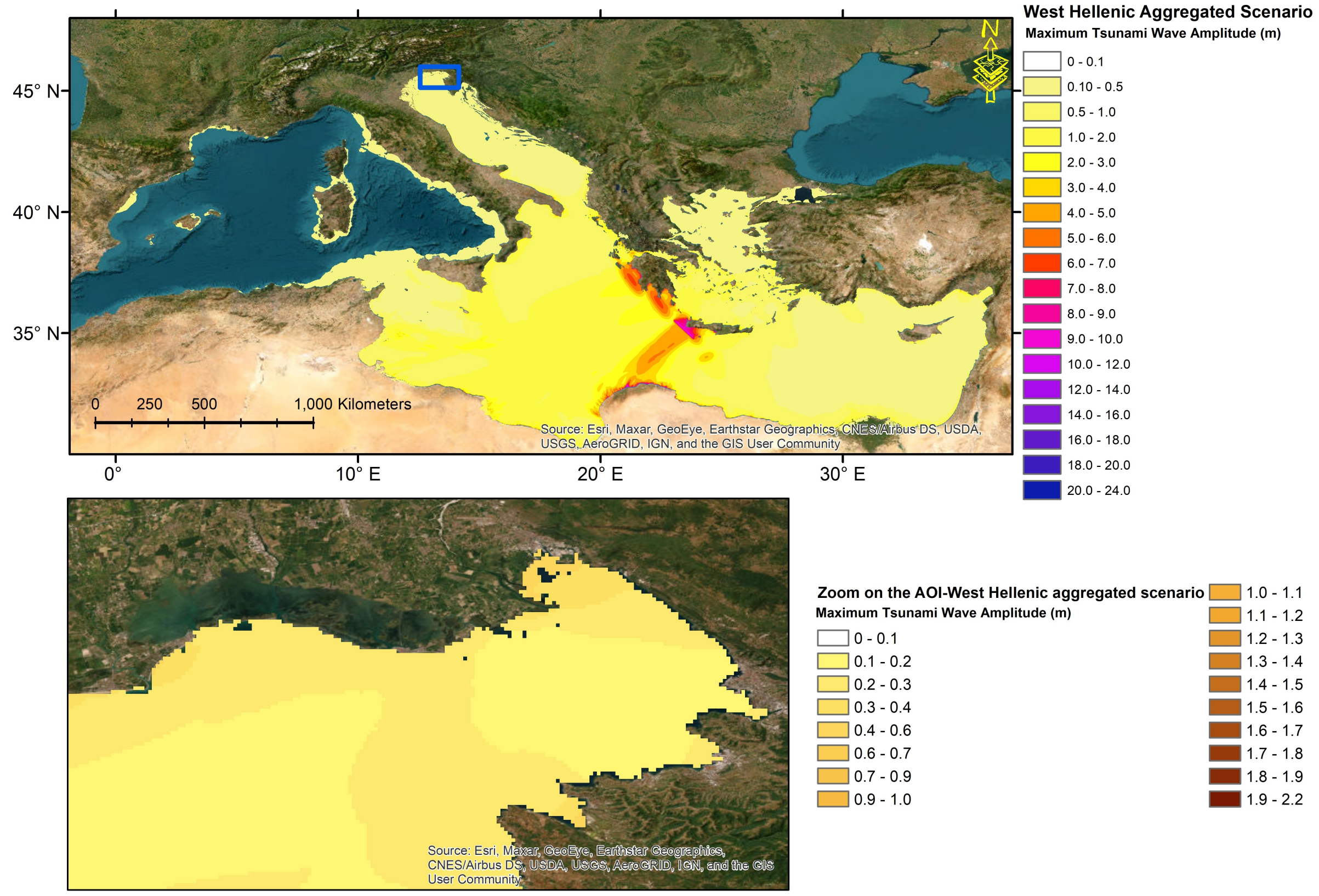

Figure 13 - Aggregated scenario from the individual scenarios computed for the West Hellenic zone at basin wide scale (top) and a zoom to the area of interest (bottom).

### 5.4 East Hellenic Arc Zone

The East Hellenic Arc is another subduction zone located in the Aegean Sea, between Greece and Turkey. Like the West Hellenic Arc, this area is also seismically active and experienced several significant earthquakes and tsunamis in the past. While the impact of a tsunami from the East Hellenic Arc on AOI is not as high as from other areas, it is still important for authorities and residents in the region to be aware of the potential related risks and take appropriate measures to mitigate them. To this end an aggregated scenario has been computed from three individual scenarios located within the East Hellenic Zone, as shown in Fig. 14. The maximum computed wave amplitude at the AOI is in the range 0.1-0.16 m (Fig. 14, bottom panel).

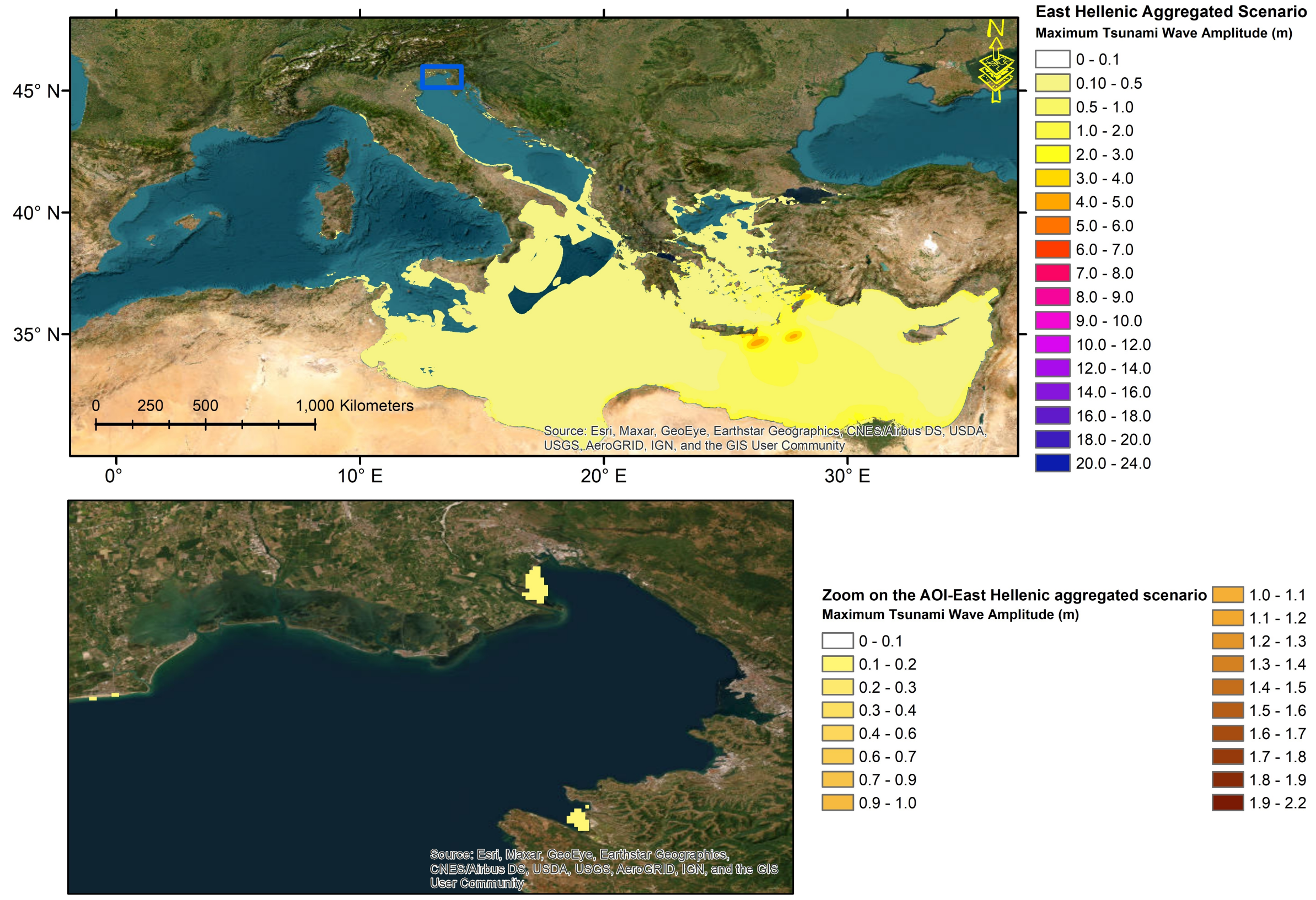

Figure 14 - Aggregated scenario from the individual scenarios computed for the East Hellenic zone at basin wide scale (top) and a zoom to the area of interest (bottom).

## 5.5 Cyprian Arc Zone

The Cyprian Arc is a subduction zone located in the eastern Mediterranean Sea, between Cyprus and the Turkish coast (Fig. 10). This area is also known to be seismically active and experienced several significant earthquakes and tsunamis in the past.

While the Adriatic Sea is relatively far from the Cyprian Arc, pre-computed scenarios may still provide useful information for emergency managers in AOI.Overall, the possible impact of a tsunami with origin in the Cyprian Arc on the coast of the AOI turns out very low and almost negligible (Fig.15). The results in Fig. 15 indicate that the tsunami associated with the largest earthquake within this zone might be insignificant (i.e. 0.03 m) in terms of hazard to the AOI.

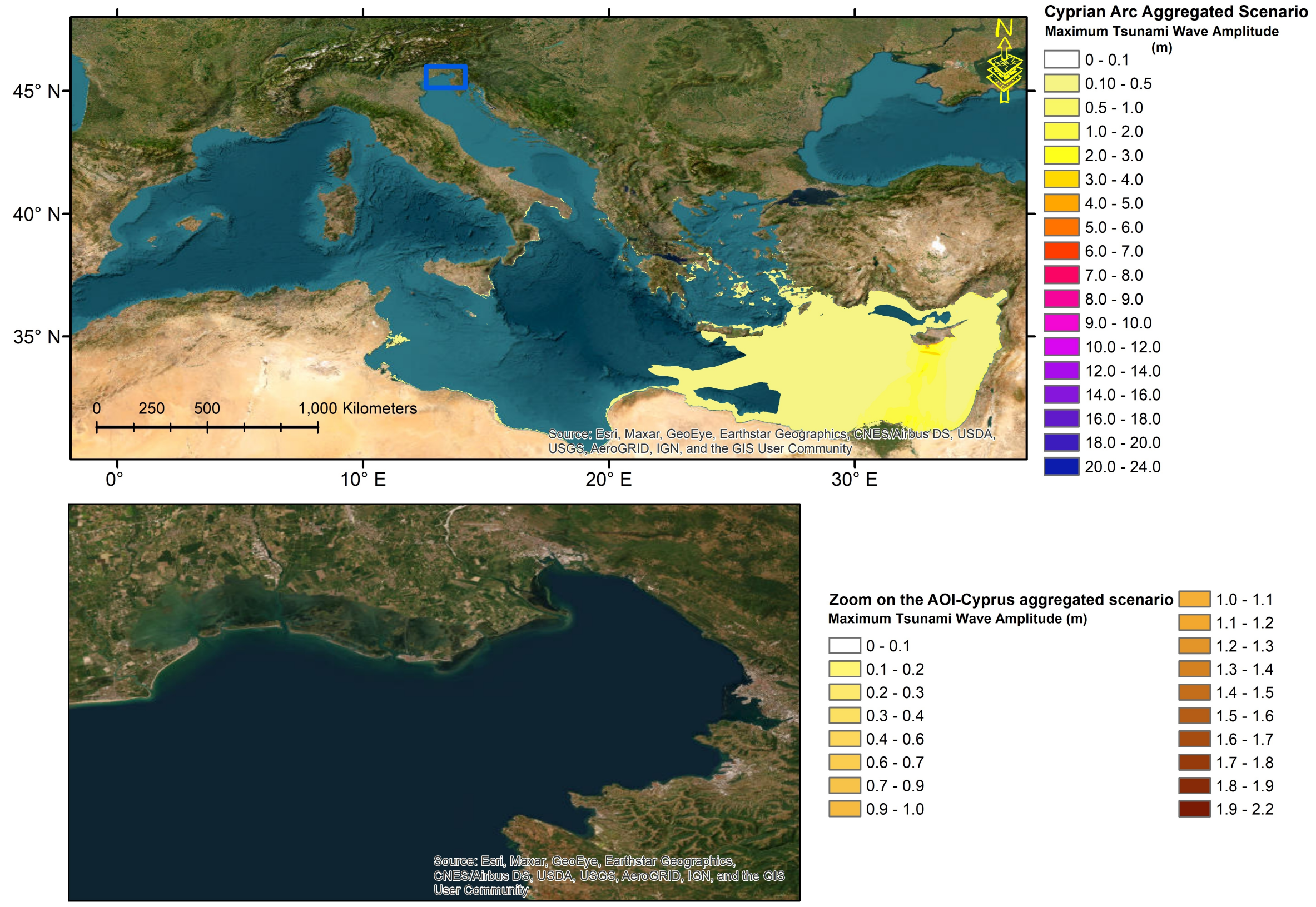

Figure 15 - Aggregated scenario from the individual scenarios computed for the Cyprian Arc Zone at basin wide scale (top) and a zoom to the area of interest (bottom).

## 6. Conclusions and future developments

In order develop tsunami hazard maps, which can be used for risk assessment and to develop evacuation plans for the areas of interests, numerical models were used to assess and predict the physical characteristics of potential tsunamis in the Adriatic Sea region.Numerical models (i.e., computer-derived simulation packages) use a grid system for the area of interest that contains information such as bathymetry, topography and surface roughness. Therefore, a numerical model can incorporate complicated geographic variations in bathymetry, topography and land uses, and can simulate different aspects of tsunami, including their variations in wave amplitude, current speed and inundation depth.

In the current work a preliminary set of tsunami scenarios was modeled, considering all potential seismogenic sources capable to generate tsunamis according to the CAT-INGV's decision matrix. For computation of tsunami hazard maps (local, regional and basin-wide), we modeled tsunamigenic sources available in recently updated the DISS-3.3,which indicated a significant increase in the $M_{max}$ of most seismogenic sources. Tsunami maximum wave amplitude, which may allow developing inundation scenarios, was computed both over a grid and for the selected POIs, including the cities of Trieste, Monfalcone, Lignano and Grado.

Adriatic-wide and regional tsunamis turn out capable to produce tsunami waveswith amplitude as large as 0.7 and 1.1 m, respectively. Local tsunamis are capable to produce tsunami wave amplitude between 2.0 and 2.2 m; their highest impact could be on the coastal zone of Trieste and Grado, while it could reach 1.2 m for Monfalcone and about 0.7 m for Lignano, as indicated in the Fig. 9 and Tables 3 and 4. The reader should keep in mind that tables presented here provide just a summary representation of the modelling results. For further details and a comprehensive description of the results, it is necessary to consider the full set of output files (in kml, grid, and dat formats), which contain complete numerical information from the simulations. Detailed analysis of these local sources, as well as the development of more detailed scenarios for the selected AOI, especially in lagoon areas, will be the main focus of future developments.

Basin-wide tsunamis from West Greece, Calabrian Arc and West Hellenic Arc should be taken into account, because tsunamigenic earthquake of $M_{max}$>7.5 turn out capable to impact the AOI at orange alert level(amplitude less than 0.5 m) and even at red alert level (up to 0.8 m for a magnitude M=8.5 earthquake in the West Hellenic Arc),as inferred from the modelling. The scenarios computed for the East Hellenic and Cyprian Arcs, instead, provide amplitudes corresponding to information level, as discussed in Section 5.

The modelling results obtained so far provide a preliminary physics based description of the tsunami hazard along Northeastern Adriatic coasts, associated with possible tsunamis generated by offshore earthquakes at different epicentral distances. Though the obtained information supplies some quantitative basis in support of the decision making process, which could be considered in the evenience a warning message is issued by the CAT-INGV tsunami alert system, there are still several aspects that need to be further investigated. In order to achieve an improved (i.e., more detailed, complete and robust) tsunami hazard assessment, the envisaged future developments include:

- In-depth revision of the high-resolution bathymetry dataset,to improve the match with observations and to remove possible artifacts (e.g., closed openings and/or channels), which could impact the propagation of tsunami waves within lakes or lagoons behind the shoreline;
- Development of inundation maps,by modelling ingression of tsunami waves, based upon very high resolution topo-bathymetry data;

- Sensitivity analysis with respect to $M_{max}$ (including the maximum credible earthquake), to account for the uncertainty on the maximum magnitude associated with the tsunamigenic sources, demonstrated by the recent revision of the DISS database;
- Detailed investigation of local tsunamigenic sources, considering different earthquake rupture styles (i.e. unilateral, bilateral), variability and directivity angles, which represent different potential scenarios that may occur in future rupture;
- Assessment of the hazard from near-offshore tsunamigenic sources, according to guidelines in the Decision Matrix, taking advantage from earlier studies in the region (e.g. Paulatto et al., 2007) and theoretical considerations (Yanovskaya et al., 2003).

Table 3 - Summary of estimated maximum tsunami wave amplitude (range) at the cities of interest, due to tsunami earthquake sources located at local, regional and Adriatic scales.

| Location | Local ≤100 km | Regional >100-≤400 km | Adriatic -wide >400 km |
|---|---|---|---|
| | Hmax (m) | Hmax (m) | Hmax (m) |
| Lignano | 0.6-0.7 | 0.2-1.1 | 0.2-0.7 |
| Grado | 2.0-2.2 | 0.2-1.1 | 0.2-0.7 |
| Monfalcone | 1.0-1.2 | 0.2-0.7 | 0.1-0.4 |
| Trieste | 1.8-2.0 | 0.2-0.8 | 0.1-0.4 |

Table 4 - List of estimated maximum tsunami wave amplitude and estimated arrival times (ETA) for the cities of interest,associated to the DISS tsunamigenic sources located at local, regional and Adriatic scales. The largest wave amplitudes estimated at each city are highlighted in grey.

| Class | DISS-ID | Name | Updated DISS-3.3 | Maximum Tsunami Wave Amplitude (m) | | | | ETA (minutes) |
|---|---|---|---|---|---|---|---|---|
| | | | $M_{max}$ | Lignano | Grado | Monfalcone | Trieste | AOI |
| **Adriatic-Wide tsunamigenic earthquake sources** | ALCS002 | Lushnje | 7.5 | <0.2 | <0.2 | <0.2 | <0.2 | 394-400 |
| | ALCS020 | Seman Coastal | 7.2 | <0.2 | <0.2 | <0.2 | <0.2 | 400-412 |
| | HRCS001 | Mljet | 7.3 | 0.5-0.6 | 0.4-0.5 | <0.2 | <0.2 | 330-340 |
| | HRCS002 | Hvar | 7.3 | >0.2 | >0.2 | <0.2 | <0.2 | 290-300 |
| | HRCS004 | Eastern Adriatic offshore - South | 6.9 | 0.3-0.4 | 0.2-0.3 | <0.2 | <0.2 | 285-295 |
| | HRCS007 | Vis-Korcula | 7.3 | 0.5-0.6 | 0.3-0.4 | 0.2-0.3 | 0.2-0.3 | 260-275 |
| | HRCS010 | Palagruza | 6.9 | 0.2-0.3 | 0.2-0.3 | 0.2-0.3 | 0.2-0.3 | 297-309 |
| | ITCS059 | S. Benedetto-Giulianova offshore | 7.5 | 0.2-0.3 | 0.2-0.3 | 0.2-0.3 | 0.2-0.3 | 230-240 |
| | ITCS074 | Shallow Gondola Fault Zone | 6.9 | 0.3-0.4 | 0.3-0.4 | 0.2-0.3 | 0.2-0.3 | 325-335 |
| | MECS001 | Montenegro offshore | 8 | 0.6-0.7 | 0.6-0.7 | 0.3-0.4 | 0.3-0.4 | 296-308 |
| | MECS007 | Budva Offshore | 7.1 | 0.3-0.4 | 0.3-0.4 | 0.2-0.3 | 0.2-0.3 | 296-308 |
| **Regional tsunamigenic earthquake sources** | HRCS008 | Dugi Otok | 7.5 | **1.0-1.1** | **0.8-0.9** | **0.4-0.5** | 0.4-0.5 | 122-135 |
| | HRCS014 | Jana-1 | 6.9 | 0.2-0.3 | 0.2-0.3 | 0.2-0.3 | 0.2-0.3 | 244-255 |
| | HRCS018 | Vis-West | 6.9 | 0.2-0.3 | 0.2-0.3 | 0.2-0.3 | 0.2-0.3 | 260-270 |
| | HRCS020 | Eastern Adriatic offshore - Central | 6.9 | 0.4-0.5 | 0.3-0.4 | 0.2-0.3 | 0.2-0.3 | 220-230 |
| | HRCS021 | Eastern Adriatic offshore - North | 6.9 | **1.0-1.1** | 0.6-0.7 | 0.3-0.4 | 0.3-0.4 | 95-110 |
| | HRCS025 | Krk | 7.2 | <0.2 | <0.2 | <0.2 | <0.2 | 168-180 |
| | ITCS004 | Castelluccio dei Sauri-Trani | 7.3 | <0.2 | <0.2 | <0.2 | <0.2 | 325-335 |
| | ITCS031 | Conero offshore | 6.6 | 0.6-0.7 | 0.6-0.7 | **0.6-0.7** | **0.6-0.7** | 168-178 |
| | ITCS039 | Riminese onshore | 7 | **1.0-1.1** | **1.0-1.1** | 0.3-0.4 | **0.7-0.8** | 155-165 |
| | ITCS070 and ITCS074 | Shallow and Deep Gondola Fault Zone | 6.9-7.2 | 0.3-0.4 | 0.3-0.4 | 0.2-0.3 | 0.2-0.3 | 325-335 |
| | ITCS106 | Pesaro mare-Cornelia | 6.5 | 0.2-0.3 | 0.2 | <0.2 | <0.2 | 162-172 |
| | ITCS108 | Clara | 6.5 | 0.2-0.3 | 0.2 | <0.2 | <0.2 | 162-172 |
| | ITCS134 | Roseto degli Abruzzi | 6.9 | <0.2 | <0.2 | <0.2 | <0.2 | 247-257 |
| | ITCS154 | Edmond | 6.8 | <0.2 | <0.2 | <0.2 | <0.2 | 248-258 |
| | ITCS155 | Daniel | 6.5 | <0.2 | <0.2 | <0.2 | <0.2 | 194-216 |
| | ITCS158 | Valeria | 6.6 | 0.2-0.3 | 0.2-0.3 | <0.2 | <0.2 | 194-216 |
| **Local tsunamigenic earthquake sources** | ITCS100 and ITCS101 | Northern Trieste Gulf<br>Southern Trieste Gulf | 6.9<br>6.5 | 0.6-0.7 | **2.0-2.2** | **1.0-1.2** | **1.8-2.0** | 15 \| 10 \| * \| * |

*The ETA is too short.

**Acknowledgments**

We are grateful to Prof. Ahmet Yalçıner and the team who developed the NAMI DANCE code for providing their user-friendly tsunami simulation package, with which the results of this study were obtained. Authors would like to acknowledge GEBCO Compilation Group (GEBCO 2020 Grid; doi:10.5285/a29c5465-b138-234d-e053-6c86abc040b9) for making bathymetry data free open access. The research benefited from financial support by Protezione Civile della Regione Autonoma Friuli-Venezia Giulia.